\documentclass[superscriptaddress,aps,pra,twocolumn,showpacs,nofootinbib,longbibliography]{revtex4-1}
\usepackage{amsmath,amssymb,amsthm}
\usepackage{easybmat}
\usepackage[colorlinks=true,citecolor=blue,urlcolor=blue]{hyperref}
\usepackage[pdftex]{graphicx}
\usepackage{times,txfonts}
\usepackage{braket}
\usepackage{color}
\usepackage{natbib}

\newcommand{\be}{\begin{equation}}
\newcommand{\ee}{\end{equation}}
\newcommand{\ba}{\begin{eqnarray}}
\newcommand{\ea}{\end{eqnarray}}
\newcommand{\ketbra}[2]{|#1\rangle \langle #2|}
\newcommand{\tr}{\operatorname{Tr}}

\newcommand{\etal}{{\it{et. al. }}}

\newtheorem{thm}{Theorem}

\newtheorem{prop}{Proposition}

\begin{document}
\title{Cost of Einstein-Podolsky-Rosen steering in the context of
extremal boxes}  
\author{Debarshi Das}
\email{debarshidas@jcbose.ac.in}
\affiliation{Centre for Astroparticle Physics and Space Science (CAPSS),
Bose Institute, Block EN, Sector V, Salt Lake, Kolkata 700 091, India}
\author{Shounak Datta}
\email{shounak.datta@bose.res.in}
\author{C.  Jebaratnam}  
\email{jebarathinam@bose.res.in} 
\author{A. S. Majumdar}
\email{archan@bose.res.in}
\affiliation{S. N. Bose National Centre for Basic Sciences, Salt Lake, Kolkata 700 098, India} 
\begin{abstract}
Einstein-Podolsky-Rosen steering is a form of quantum nonlocality which is  weaker than
Bell nonlocality, but stronger than entanglement. 
Here we present a method to check Einstein-Podolsky-Rosen steering in the scenario
where the steering party performs two black-box measurements and the trusted party performs 
projective qubit measurements corresponding to two arbitrary mutually unbiased bases. 
This method is based on decomposing the  
measurement correlations in terms of extremal boxes of the steering scenario. 
In this context, we propose a measure
of steerability called steering cost. 
We show that our steering cost is a convex steering monotone.
We illustrate our method to check steerability with two families of measurement correlations
and find out their steering cost.
\end{abstract}

\maketitle 

\section{Introduction}

Quantum entanglement admits stronger than classical correlations which can lead to 
quantum nonlocality. 
Local quantum measurements on a composite system lead to nonlocality 
if the statistics of the measurement outcomes  cannot be  
explained by a local hidden variable (LHV) model \cite{Bel64,BCP+14}. Such a nonclassical feature 
of quantum correlations termed as Bell nonlocality can be used to certify the presence of entanglement
in a device-independent way and it finds applications in device-independent
quantum information processing \cite{BCP+14}. 

Quantum steering is a form of quantum nonlocality 
which was first noticed by Schrodinger \cite{Schrodinger} in the context of the famous
Einstein-Podolsky-Rosen (EPR) paradox \cite{EPR35}.
EPR steering arises in the scenario where local quantum measurements on one part of a bipartite system
are used to prepare different ensembles for the other part. This scenario 
demonstrates EPR steering if these ensembles cannot be explained by a local hidden state (LHS)
model \cite{WJD07}. The demonstration of the EPR paradox was first proposed by Reid \cite{Rei89} based 
on the Heisenberg uncertainty relation. Using tighter uncertainty relations such as entropic
ones, corresponding entropic steering criteria have been subsequently proposed \cite{WSG+11}, leading to the demonstration of
steering for more categories of states \cite{CPM+14}.
Oppenheim and Wehner \cite{OW10} introduced fine-grained uncertainty relations that provide
 a direct way of linking uncertainty with nonlocality.
In Ref \cite{PKM14}, Pramanik \etal  have derived steering inequalities based 
on fine-grained uncertainty relations, an approach that has been later extended for continuous variables too \cite{CPM15}.

It is well-known that EPR steering lies in between entanglement and Bell nonlocality:
quantum states that demonstrate Bell nonlocality form a subset of EPR steerable states which,
in turn, form a subset of entangled states \cite{WJD07,QVC+15}. The operational definition of
EPR steering is that it certifies 
the presence of entanglement in a one-sided device-independent way in which the measurement
device at only one of the two sides is fully trusted \cite{JWD07}.
Steering inequalities  
which are analogous to Bell inequalities have been derived to rule out LHS description
for the steering scenarios \cite{CJW+09,ZHC16}. 
Recently, it has been demonstrated that 
violation of a steering inequality is necessary for one-sided device-independent quantum key distribution \cite{BCW+12}.
EPR steering admits
an asymmetric formulation: there exist entangled states which are one-way steerable, i.e.,
demonstrate steerability from one observer to the other observer but not vice-versa \cite{CYW+13,BVQ+14}.   
Various other steering criteria have also been proposed such as
all versus nothing proof of EPR steering \cite{CYW+13} and hierarchy of steering criteria based on moments \cite{KSC+15}.

Motivated by the question of how much a steering scenario demonstrates steerability,
a measure of steering called steering weight was defined in Ref. \cite{SNC14}.
Quantitative characterization of steering has started receiving  attention recently \cite{GA15,CBL+16}.
In Ref. \cite{GA15}, Gallego and Aolita (GA) have developed the resource theory of steering.
GA have observed that in the steering theory,  local operations assisted by one-way 
classical communications ($1$W-LOCCs) from
the trusted side to the black-box side are allowed operations.
With $1$W-LOCCs as free operations of steering, GA have introduced a set of postulates that a bona
fide quantifier of steering should fulfill. Those functions that satisfy these postulates are called convex steering
monotones. GA have proved that the first proposed measure of steering, i.e., steering weight is a convex steering monotone.

In the case of the Bell scenario with a finite number of settings per party and a finite number of
outcomes per setting, it is well-known that the set of correlations that have a LHV model
forms a convex polytope \cite{Fin82,PR94,BLM+05}. The nontrivial facet inequalities of this polytope are called 
Bell inequalities. For a given Bell scenario, a correlation has a LHV model iff (if and only if)
it satisfies all the Bell inequalities.
In Ref. \cite{CFF+15}, Cavalcanti, Foster, Fuwa and Wiseman (CFFW) have considered an analogous characterization 
of EPR steering. Steering can also be understood as a failure of a hybrid 
local hidden variable-local hidden state (LHV-LHS) model to produce the correlations between the black-box side 
and the trusted side. In Ref. \cite{CFF+15}, CFFW have shown that any 
LHV-LHS model can be written as a convex mixture of the extremal points 
of the unsteerable set.

In this work, we present a method to check EPR steering in the context of extremal points
of the following steering scenario: Alice performs two black-box measurements and Bob performs projective qubit measurements corresponding to any two mutually unbiased bases (MUBs). 
This method provides a simple way to check the existence of a LHV-LHS model
for the  measurement correlations arising from the above steering scenario.
Based on this formulation, we propose a measure of
steerability which we call steering cost.   
We show  that our steering cost is a convex steering monotone.
We illustrate our method to check steerability with two families of measurement correlations
and we find out the steering cost of these two families.  Steering cost is also compared with another measure of steering, called ``steering weight" \cite{SNC14}. The advantage in experimental determination of steering cost over that of steering weight is also discussed.

The organization of the paper is as follows. In Sec. \ref{prl}, we review the polytope of 
nonsignaling boxes which we use to provide a criterion for EPR steering and discuss some basic notions in EPR steering.
In Sec. \ref{qepr}, we present our quantifier of steering and we apply our method to check steerability of
two families of measurement correlations.  Comparison of steering cost with steering weight is presented in Sec. \ref{scsw}.
In Sec. \ref{conc}, we present 
our concluding remarks. 

\section{Preliminaries} \label{prl}

\subsection{Bell nonlocality} 

Consider the Bell scenario where two spatially separated parties, Alice and Bob,
share a bipartite black box. Let us denote the inputs on  Alice's and Bob's sides by $x$ and $y$,
respectively, and the outputs by $a$ and $b$. The given Bell scenario is 
characterized by the set of joint probabilities,  $P(ab|xy):=\{p(a b|x y)\}_{a,x,b,y}$,
which is called correlation or box (also denoted by $P$). 
A correlation  $P$ is Bell nonlocal
if it cannot be reproduced by a LHV model, i.e., 
\be
p(ab|xy)=\sum_\lambda p(\lambda) p(a|x,\lambda)p(b|y,\lambda) \hspace{0.3cm} \forall a,b,x,y,
\ee
where $\lambda$ denotes shared randomness which occurs with probability $p(\lambda)$; each $p(a|x,\lambda)$ and $p(b|y,\lambda)$ are conditional probabilities.

In the case of two-binary-inputs and two-binary-outputs per side,
the set of nonsignaling boxes forms an $8$ dimensional
convex  polytope with  $24$ extremal  boxes \cite{BLM+05}, the  $8$ Popescu-Rohrlich
(PR) boxes \cite{PR94}: 
\begin{align}
&P_{PR}^{\alpha\beta\gamma}(ab|xy) \nonumber
\\&=\left\{
\begin{array}{lr}
\frac{1}{2}, & a\oplus b=x\cdot y \oplus \alpha x\oplus \beta y \oplus \gamma\\ 
0 , & \text{otherwise}\\
\end{array}
\right. \label{NLV}
\end{align}
and $16$ local-deterministic boxes:
\begin{equation}
P_D^{\alpha\beta\gamma\epsilon}(ab|xy)=\left\{
\begin{array}{lr}
1, & a=\alpha x\oplus \beta\\
   & b=\gamma y\oplus \epsilon \\
0 , & \text{otherwise}.\\
\end{array}
\right.   
\label{eq:locdet}
\end{equation}   
Here, $\alpha,\beta,\gamma,\epsilon\in  \{0,1\}$ and  $\oplus$ denotes
addition modulo  $2$. All the deterministic boxes as defined above
can be written as the product of marginals corresponding to Alice and Bob,
i.e., $P_D^{\alpha\beta\gamma\epsilon}(ab|xy)=P^{\alpha\beta}_D(a|x)P^{\gamma\epsilon}_D(b|y)$,
with the  deterministic box on Alice's side given by,
\begin{equation}
P_D^{\alpha\beta}(a|x)=\left\{
\begin{array}{lr}
1, & a=\alpha x\oplus \beta\\
0 , & \text{otherwise}\\
\end{array}
\right. 
\label{}
\end{equation} 
and the  deterministic box on Bob's side given by,
\begin{equation}
P_D^{\gamma\epsilon}(b|y)=\left\{
\begin{array}{lr}
1, & b=\gamma x\oplus \epsilon\\
0 , & \text{otherwise}.\\
\end{array}
\right.   
\label{}
\end{equation} 

 The $8$ PR boxes are equivalent under ``local reversible operations" (LRO). Similarly, the $16$ local-deterministic boxes are equivalent under LRO. By using LRO Alice and Bob can convert any PR box into any other PR box, or any local-deterministic box into any other local-deterministic box. LRO is designed  \cite{BLM+05} as follows: Alice may relabel her inputs: $x \rightarrow x \oplus 1$, and she may relabel her outputs (conditionally on the input) : $a \rightarrow a \oplus \alpha x \oplus \beta$; Bob can perform similar operations.

The set of boxes which have a LHV model
forms a subpolytope of the full nonsignaling polytope whose extremal boxes are the 
local-deterministic boxes. A box with two-binary-inputs-two-binary-outputs
is local iff it satisfies a Bell--Clauser-Horne-Shimony-Holt (CHSH) inequality \cite{CHS+69} and  its 
permutations \cite{Fin82}  which are given by,
\begin{align}
&\mathcal{B}_{\alpha\beta\gamma} := (-1)^\gamma\braket{A_0B_0}+(-1)^{\beta \oplus \gamma}\braket{A_0B_1}\nonumber\\
&+(-1)^{\alpha \oplus \gamma}\braket{A_1B_0}+(-1)^{\alpha \oplus \beta \oplus \gamma \oplus 1} \braket{A_1B_1}\le2,
\label{BCHSH}
\end{align}
where $\braket{A_xB_y}=\sum_{ab}(-1)^{a \oplus b} P(ab|xy)$.
The above inequalities form the facet inequalities for the local polytope formed 
by the extremal points given in Eq. (\ref{eq:locdet}).

Nonlocal cost is a measure of nonlocality \cite{BCS+11} which is based on the 
Elitzur-Popescu-Rohrlich decomposition \cite{EPR92}. In this approach, a given
box $P(ab|xy)$ is decomposed into a nonlocal part  and a local part, i.e.,
\be
P(ab|xy)=p_{NL}P_{NL}(ab|xy)+(1-p_{NL}) P_{L}(ab|xy), \label{nlcost}
\ee
where $P_{NL}(ab|xy)$  (or, simply, $P_{NL}$) is a nonsignaling box and $P_{L}(ab|xy)$  (or, simply, $P_{L}$) is a local box; $0 \leq p_{NL} \leq 1$.
The nonlocal cost of the box $P(ab|xy)$, denoted $C_{NL}(P)$, is obtained by minimizing
the weight of the nonlocal part over all possible decompositions of the 
form (\ref{nlcost}), i.e.,
\be
C_{NL}(P):=\min_{decompositions}p_{NL}.
\ee
Here, $0 \leq C_{NL}(P) \leq 1$. It turns out that, for the optimal decomposition, the nonlocal part $P_{NL}(ab|xy)$ has the maximal 
nonlocal cost, i.e., $C_{NL}(P_{NL})=1$ 
since it is an extremal nonlocal box. An extremal nonlocal box in a given Bell scenario
cannot be decomposed as a convex mixture of the other boxes in that given Bell scenario and violates a Bell inequality maximally \cite{BLM+05}.
In the case of two-binary-inputs and two-binary-outputs per side, for the optimal decomposition,
the nonlocal part $P_{NL}(ab|xy)$ is one of the PR-boxes given in Eq. (\ref{NLV}). 

\subsection{EPR steering} 

Consider a steering scenario where Alice and Bob share an unknown quantum system
described by $\rho_{AB}\in \mathcal{B}(\mathcal{H}_A \otimes \mathcal{H}_B)$, with Alice 
performing a set of black-box measurements  and the Hilbert-space dimension of Bob's 
subsystem is known. Such a scenario is called one-sided device-independent since
Alice's measurement operators ${\bf{M}}_A:=\{M_{a|x}\}_{a,x}$  are unknown. The steering scenario 
is completely characterized by an assemblage \cite{Pus13} ${\pmb{\sigma}}:=\{\sigma_{a|x}\}_{a,x}$
which is the set of unnormalized conditional states on Bob's
side. Each element
in the assemblage ${\pmb{\sigma}}$ 
is given by $\sigma_{a|x}=p(a|x)\rho_{a|x}$,  where $p(a|x)$ is the conditional probability of getting the outcome of Alice's measurement  and
$\rho_{a|x}$ is the normalized conditional state on Bob's side.
Quantum theory predicts the assemblage as follows:
\be
\sigma_{a|x}=\tr_A ( M_{a|x} \otimes \openone \rho_{AB}) \hspace{0.5cm} \forall \sigma_{a|x} \in {\pmb{\sigma}}  \label{as}.
\ee
Let $\Sigma^{S}$ denote the set of all valid assemblages. 

In the above scenario, Alice demonstrates steerability to Bob
if the assemblage does not have a local hidden state (LHS) model, i.e., if for all $a$, $x$, there
is no decomposition of $\sigma_{a|x}$ in the form,
\be
\sigma_{a|x}=\sum_\lambda p(\lambda) p(a|x,\lambda) \rho_\lambda,
\ee
where $\lambda$ denotes classical random variable which occurs with probability 
$p(\lambda)$; $\rho_{\lambda}$
are called local hidden states which satisfy $\rho_\lambda\ge0$ and
$\tr\rho_\lambda=1$. Let $\Sigma^{US}$ denote the set of all unsteerable assemblages. 
Any element in the given assemblage ${\pmb{\sigma}} \in \Sigma^{US}$
 can be decomposed
in terms of deterministic distributions  as follows:
\be
\sigma_{a|x}=\sum_\chi D(a|x,\chi) \sigma_\chi,
\ee
where $D(a|x,\chi):=\delta_{a,f(x,\chi)}$ is the single-partite
extremal conditional probability for Alice determined by the variable 
$\chi$ through the function $f(x,\chi)$ and $\sigma_\chi$ satisfy
$\sigma_\chi\ge0$ and $\sum_{\chi} \tr (\sigma_\chi)=1$. For a given scenario, 
the above decomposition has been used to define semi-definite 
programming to check steerability \cite{CS17}.

Suppose Bob performs a set of projective measurements ${\bf{\Pi}}_B:=\{\Pi_{b|y}\}_{b,y}$ 
 on ${\pmb{\sigma}}$. Then the
scenario is characterized by the set of measurement correlations which is a box
shared by Alice and Bob, $P(ab|xy)$=$\Big\{ \tr\big[\Pi_{b|y} \sigma_{a|x} \big] \Big\}_{a,x,b,y}$.
If the box $P(ab|xy)$ detects steerability from Alice to Bob, 
then it does not have a decomposition as follows:
\be
p(ab|xy)= \sum_\lambda p(\lambda) p(a|x,\lambda) p(b|y, \rho_\lambda) \hspace{0.3cm} \forall a,x,b,y, \label{LHV-LHS}
\ee
where  $p(b|y, \rho_\lambda)=\tr(\Pi_{b|y}\rho_\lambda)$, which arises from some local hidden state $\rho_\lambda$.
The above decomposition is called a LHV-LHS model.
Let us denote the set of all correlations that belongs to the given steering scenario $\mathcal{N}_{\Sigma^{S}}$. 
The set of correlations that have a LHV-LHS model denoted by $\mathcal{L}_{\Sigma^{US}}$ forms a convex 
subset of $\mathcal{N}_{\Sigma^{S}}$ \cite{CJW+09}, which we call unsteerable set.
In particular, any LHV-LHS model can be decomposed in terms of the extremal points 
of $\mathcal{L}_{\Sigma^{US}}$ \cite{CFF+15}.
That is we can simplify the decomposition (\ref{LHV-LHS}) as follows:
\be
p(ab|xy)=\sum_{\chi,\zeta} p(\chi,\zeta)D(a|x,\chi)\braket{\psi_\zeta|\Pi_{b|y}|\psi_\zeta} \hspace{0.2cm} \forall a,x,b,y, \label{DLHS}
\ee
with $D(a|x,\chi)=\delta_{a,f(x,\chi)}$. Here, $\chi$ are the variables which determine all values 
of Alice's observables $A_x$ through the function $f(x,\chi)$ and $\zeta$ determines a pure state $\ket{\psi_\zeta}$
for Bob.

\section{Quantifying EPR steering}\label{qepr}

Analogous to nonlocal cost, we now define steering cost of a box  $P(ab|xy) \in \mathcal{N}_{\Sigma^{S}}$. 
First, the given box $P(ab|xy)$ is decomposed in a convex mixture of a steerable part
and an unsteerable part, i.e., 
\be
P(ab|xy)=p_{S}P_{S}(ab|xy)+(1-p_{S})P_{US}(ab|xy), \label{steercost}
\ee
where $P_{S}(ab|xy)$ (or, simply, $P_{S}$) is a steerable box and $P_{US}(ab|xy)$  (or, simply, $P_{US}$) is an unsteerable box; $0 \leq p_S \leq 1$.
Second, the weight of the steering  part minimized over all possible decompositions of 
the  form (\ref{steercost}) gives 
the steering cost of the box $P(ab|xy)$ denoted by $C_{steer}(P)$, i.e.,
\be
C_{steer}(P):=\min_{decompositions}p_{S}.
\ee
Here, $0 \leq C_{steer}(P) \leq 1$ (since, $0 \leq p_S \leq 1$). It follows that, for the optimal decomposition, the steerable part $P_{S}(ab|xy)$ has the maximal 
steering cost, i.e.,  $C_{steer}(P_S)=1$ since it is an extremal steerable box. An extremal steerable box $P^{Ext}_{S}(ab|xy)\in \mathcal{N}_{\Sigma^{S}}$
cannot be decomposed as a convex mixture of the other boxes in the set $\mathcal{N}_{\Sigma^{S}}$ and violates
a steering inequality in the given steering scenario maximally.\\

We will now demonstrate that the steering cost $C_{steer}(P)$ is a proper quantifier of steering, i.e., it is
a convex steering monotone \cite{GA15}. For this purpose, we introduce the following notations.  
A box $P(ab|xy)$ which is obtained 
by Bob performing projective measurements ${\bf{\Pi}}_B$ on an assemblage ${\pmb{\sigma}}$ is  denoted by 
$P[{\pmb{\sigma}}]$. Here, $P[{\pmb{\sigma}}]:=P(ab|xy)=\Big\{ \tr\big[\Pi_{b|y} \sigma_{a|x} \big] \Big\}_{a,x,b,y}$. 
Consider the situation in which deterministic {\it one-way local operations and classical communications} ($1$W-LOCCs) \cite{GA15}  
 occur 
from Bob to Alice before Bob performs measurements
on the assemblage. Following Ref. \cite{HLL16}, we define the deterministic $1$W-LOCC as
a completely positive trace preserving (CPTP) map $\mathcal{M}$ that take an assemblage
${\pmb{\sigma}}$ into a final assemblage $\mathcal{M}({\pmb{\sigma}})$, where
\be
\mathcal{M}({\pmb{\sigma}})=\sum_{\omega} \mathcal{M}_{\omega} ({\pmb{\sigma}}) 
:=\sum_\omega \mathcal{K}_{\omega} \mathcal{W}_{\omega} ({\pmb{\sigma}}) \mathcal{K}_{\omega}^{\dagger},
\ee
with $\mathcal{W}_{\omega}$ being a deterministic wiring map which transforms one assemblage ${\pmb{\sigma}}$ = $\lbrace \sigma_{a \vert x} \rbrace_{a,x}$ to another assemblage ${\pmb{\tilde{\sigma}}}$ = $\lbrace \tilde{\sigma}_{a^{\prime} \vert x^{\prime}} \rbrace_{a^{\prime},x^{\prime}}$ having different setting $x^{\prime}$ and outcome $a^{\prime}$ at Alice's side in the following way:
\begin{align}
\label{map11}
[\mathcal{W}_{\omega}({\pmb{\sigma}})]_{x^{\prime}} &:= \tilde{\sigma}_{a^{\prime}\vert x^{\prime}} \nonumber\\
& = \sum_{a,x} p(x|x^{\prime},\omega) p(a^{\prime}|x^{\prime},a,x,\omega) \sigma_{a|x} \nonumber\\
&\forall a^{\prime}, x^{\prime}.
\end{align}
Define
\be
\mathcal{D}_{\omega}({\pmb{\sigma}}) 
:= \frac{\mathcal{M}_{\omega} ({\pmb{\sigma}})}{\tr[\mathcal{M}_{\omega}({\pmb{\sigma}})]}, \nonumber
\ee
which is the set of normalized conditional states arising from the 
action of a subchannel $\mathcal{M}_{\omega}$, labeled by $\omega$, of the CPTP map $\mathcal{M}$  on the assemblage ${\pmb{\sigma}}$ at Bob's end. Here, $\mathcal{T}(\omega) := \tr[\mathcal{M}_{\omega} ({\pmb{\sigma}})]$ is the probability of transmitting the assemblage ${\pmb{\sigma}}$ through the $\omega$th subchannel of $\mathcal{M}$; and $\sum_{\omega} \mathcal{T}(\omega) \leq 1$.
Let us denote $\mathcal{D}_{\omega}({\pmb{\sigma}})$ := $\{[\mathcal{D}_{\omega}({\pmb{\sigma}})]_{a'|x'}\}_{a^{\prime}, x^{\prime}}$, where the normalized state $[\mathcal{D}_{\omega}({\pmb{\sigma}})]_{a'|x'}$ 
denotes an element of $\mathcal{D}_{\omega}({\pmb{\sigma}})$. Hence, we can define $P[\mathcal{D}_{\omega}({\pmb{\sigma}})]$
which is a box arising from any  valid assemblage (steerable or unsteerable) 
${\pmb{\sigma}} \in \Sigma^{S}$ after the action of a map $\mathcal{M}_{\omega}$ as follows:
\begin{align}\label{Dbox}
&P[\mathcal{D}_{\omega}({\pmb{\sigma}})] = P(a^{\prime} b|x^{\prime} y) \nonumber \\
&:= \Big\{p(a^{\prime}|x^{\prime}) \tr[\Pi_{b|y} [\mathcal{D}_{\omega}({\pmb{\sigma}})]_{a'|x'}]\Big\}_{a^{\prime}, x^{\prime}, b , y}, 
\end{align}
where $p(a^{\prime}|x^{\prime})$ is the conditional probability of obtaining the outcome $a^{\prime}$, when Alice performed the measurement $x^{\prime}$, and is given by 
\be
p(a'|x') = \sum_{a,x} p(x|x', \omega) p(a'|x', a, x, \omega) p(a|x).\nonumber
\ee
This can be obtained from Eq. (\ref{map11}), expressing the elements of the  assemblages $\sigma_{a|x}$ and $\tilde{\sigma}_{a'|x'}$ at Bob's side as $p(a|x) \rho_{a|x}$ and $p(a'|x') \tilde{\rho}_{a'|x'}$ respectively (where $p(a|x)$ and $p(a'|x')$ are conditional probabilities and $\rho_{a|x}$ and $\tilde{\rho}_{a'|x'}$ 
are normalized states at Bob's side) and taking trace on both side of the equation.

With the above notations, we now proceed to show that $C_{steer}(P[{\pmb{\sigma}}])$  
satisfies the following two properties:
\begin{enumerate}
\item  $C_{steer}(P[{\pmb{\sigma}}])$ does not increase on average under deterministic 
1W-LOCCs, i.e.,
\begin{align} \label{ADLOCC}
&\sum_{\omega} \mathcal{T}(\omega) C_{steer}(P[\mathcal{D}_{\omega}({\pmb{\sigma}})]) \nonumber \\
&\leq C_{steer}(P[{\pmb{\sigma}}]), \hspace{0.3cm} \forall {\pmb{\sigma}} \in \Sigma^{S}.
\end{align}

\begin{proof}
Let us consider the following decomposition of an arbitrary assemblage ${\pmb{\sigma}} :=\{\sigma_{a|x}\}_{a,x} \in \Sigma^{S}$:
\be
\sigma_{a|x}=p_S  \sigma^S_{a \vert x} + (1-p_S) \sigma^{US}_{a \vert x} \hspace{0.2cm} \forall a,x, \label{asde}
\ee
where $\sigma^S_{a \vert x}$ is an element of an   assemblage  ${\pmb{\sigma}}^S$ having steerability and $\sigma^{US}_{a \vert x}$ is an element of an unsteerable assemblage 
${\pmb{\sigma}}^{US}$.
Now, one can write,
\begin{align}
\label{ee}
\tr\big[\Pi_{b|y} \sigma_{a|x} \big]  = &p_S \tr\big[\Pi_{b|y} \sigma^S_{a|x} \big] + (1-p_S) \tr\big[\Pi_{b|y} \sigma^{US}_{a|x} \big] \nonumber\\
& \forall a,x,b,y.
\end{align}
Hence, for the box $P[{\pmb{\sigma}}]$ arising from the assemblage ${\pmb{\sigma}}$, one can write the following decomposition: 
\begin{align}
P[{\pmb{\sigma}}]  &= p_S P_S[{\pmb{\sigma}}^S] + (1-p_S) P_{US}[{\pmb{\sigma}}^{US}]. \label{bef}
\end{align}
Here, $P_S[{\pmb{\sigma}}^S]$  is a steerable box, produced from the steerable assemblage ${\pmb{\sigma}}^S$ and 
$P_S[{\pmb{\sigma^{US}}}]$ is an unsteerable box, produced from the unsteerable assemblage ${\pmb{\sigma^{US}}}$. 
The steering cost of the box $P[{\pmb{\sigma}}]$, i.e., $C_{steer}(P[{\pmb{\sigma}}])$ is obtained by minimizing $p_S$ in Eq. (\ref{bef}) over all such possible decompositions.
Let the decomposition (\ref{bef}) denote the optimal decomposition, i.e., 
$p_{S}=C_{steer}(P[{\pmb{\sigma}}])$.\\
Now consider the set of normalized states $\mathcal{D}_{\omega}({\pmb{\sigma}})$,
where $\mathcal{D}_{\omega}$ has been applied on the assemblage ${\pmb{\sigma}}$ 
producing the box $P[{\pmb{\sigma}}]$ with the optimal decomposition given by Eq. (\ref{bef}) with $p_{S}=C_{steer}(P[{\pmb{\sigma}}])$.
From Eq. (\ref{asde}), we have,
\ba
\label{eq1}
\mathcal{D}_{\omega}({\pmb{\sigma}}) &=& 
\mathcal{D}_{\omega}\Big(p_S {\pmb{\sigma}}^S + (1-p_S) {\pmb{\sigma}}^{US} \Big) \nonumber\\
&=& \frac{\mathcal{M}_{\omega}\Big(p_S {\pmb{\sigma}}^S + (1-p_S) {\pmb{\sigma}}^{US} \Big)}{\tr[\mathcal{M}_{\omega}({\pmb{\sigma}})]},
\ea
where
\begin{align}
\label{eq2}
&\mathcal{M}_{\omega}\Big(p_S {\pmb{\sigma}}^S + (1-p_S) {\pmb{\sigma}}^{US}\Big) \nonumber\\
&= \mathcal{K}_{\omega} \mathcal{W}_{\omega} \Big(p_S {\pmb{\sigma}}^S + (1-p_S) {\pmb{\sigma}}^{US} \Big) \mathcal{K}_{\omega}^{\dagger}. 
\end{align}
Now, consider the assemblage,
\begin{equation}
\label{eee}
{\pmb{\tilde{\sigma}}} = \{ \tilde{\sigma}_{a^{\prime}|x^{\prime}}\}_{a^{\prime}, x^{\prime}} = \mathcal{W}_{\omega} \Big(p_S {\pmb{\sigma}}^S + (1-p_S) {\pmb{\sigma}}^{US} \Big).
\end{equation}
From Eq. (\ref{map11}), it follows that each element in the above assemblage ${\pmb{\tilde{\sigma}}}$ 
has the following decomposition:
\begin{align}
\tilde{\sigma}_{a^{\prime}|x^{\prime}} &=  \sum_{a,x} p(x|x^{\prime},\omega) p(a^{\prime}|x^{\prime},a,x,\omega)   \Big(p_S \sigma^S_{a \vert x} + (1-p_S) \sigma^{US}_{a \vert x}\Big)\nonumber\\ 
&= p_S \sum_{a,x} p(x|x^{\prime},\omega) p(a^{\prime}|x^{\prime},a,x,\omega)    \sigma^S_{a \vert x} \nonumber\\ 
&+ (1-p_{S})\sum_{a,x} p(x|x^{\prime},\omega) p(a^{\prime}|x^{\prime},a,x,\omega)    \sigma^{US}_{a \vert x} \nonumber\\ 
& \hspace{0.5cm} \forall a^{\prime}, x^{\prime},
\end{align}
which implies that
\begin{align}
\label{eq3}
& \mathcal{W}_{\omega} \Big(p_S {\pmb{\sigma}}^S + (1-p_S) {\pmb{\sigma}}^{US} \Big) \nonumber\\
&= p_S \mathcal{W}_{\omega} ({\pmb{\sigma}}^S) + (1-p_S)  \mathcal{W}_{\omega} ({\pmb{\sigma}}^{US}).
\end{align}
Hence, from Eqs. (\ref{eq2}) and (\ref{eq3}), we obtain
\begin{align}
\label{eq4}
&\mathcal{M}_{\omega}\Big(p_S {\pmb{\sigma}}^S + (1-p_S) {\pmb{\sigma}}^{US} \Big) \nonumber\\ 
&= \mathcal{K}_{\omega} \Big[ p_S \mathcal{W}_{\omega} ({\pmb{\sigma}}^S) + (1-p_S)  \mathcal{W}_{\omega} ({\pmb{\sigma}}^{US}) \Big] \mathcal{K}_{\omega}^{\dagger} \nonumber\\ 
&= p_S \mathcal{K}_{\omega} \Big[ \mathcal{W}_{\omega} ({\pmb{\sigma}}^S) \Big] \mathcal{K}_{\omega}^{\dagger} + (1-p_S) \mathcal{K}_{\omega} \Big[ \mathcal{W}_{\omega} ({\pmb{\sigma}}^{US}) \Big] \mathcal{K}_{\omega}^{\dagger}\nonumber\\ 
&=p_S \mathcal{M}_{\omega}( {\pmb{\sigma}}^S) + (1-p_S) \mathcal{M}_{\omega}({\pmb{\sigma}}^{US})
\end{align}
Now, from Eqs. (\ref{eq1}) and (\ref{eq4}), we obtain
\begin{align}
\label{eq5}
\mathcal{D}_{\omega}({\pmb{\sigma}}) &= 
 \frac{p_S \mathcal{M}_{\omega}( {\pmb{\sigma}}^S) + (1-p_S) \mathcal{M}_{\omega}({\pmb{\sigma}}^{US})}{\tr[\mathcal{M}_{\omega}({\pmb{\sigma}})]} \nonumber\\
&= p_S \mathcal{D}_{\omega}({\pmb{\sigma}}^S) \frac{\tr[\mathcal{M}_{\omega}({\pmb{\sigma}}^S)]}{\tr[\mathcal{M}_{\omega}({\pmb{\sigma}})]}  \hspace{0.1cm}+  \nonumber\\
& (1-p_S) \mathcal{D}_{\omega}({\pmb{\sigma}}^{US}) \frac{\tr[\mathcal{M}_{\omega}({\pmb{\sigma}}^{US})]}{\tr[\mathcal{M}_{\omega}({\pmb{\sigma}})]},
\end{align}
which implies that each element of $\mathcal{D}_{\omega}({\pmb{\sigma}})$ has the following decomposition: 
\begin{align}
\label{eq6}
[\mathcal{D}_{\omega}({\pmb{\sigma}})]_{a'|x'} 
&= p_S [\mathcal{D}_{\omega}({\pmb{\sigma}}^S)]_{a'|x'} \frac{\tr[\mathcal{M}_{\omega}({\pmb{\sigma}}^S)]}{\tr[\mathcal{M}_{\omega}({\pmb{\sigma}})]}  \hspace{0.1cm}+  \nonumber\\
& (1-p_S) [\mathcal{D}_{\omega}({\pmb{\sigma}}^{US})]_{a'|x'} \frac{\tr[\mathcal{M}_{\omega}({\pmb{\sigma}}^{US})]}{\tr[\mathcal{M}_{\omega}({\pmb{\sigma}})]} \nonumber\\
&\forall a^{\prime}, x^{\prime}.
\end{align} 
From Eq. (\ref{eq6}), one can write,
\begin{align}
\label{eq7}
& p(a^{\prime}|x^{\prime})\tr\Big[\Pi_{b|y} [\mathcal{D}_{\omega}({\pmb{\sigma}})]_{a'|x'}\Big] \nonumber \\
&= p_S \frac{\tr[\mathcal{M}_{\omega}({\pmb{\sigma}}^S)]}{\tr[\mathcal{M}_{\omega}({\pmb{\sigma}})]} p(a^{\prime}|x^{\prime})\tr\Big[\Pi_{b|y} [\mathcal{D}_{\omega}({\pmb{\sigma}}^S)]_{a'|x'} \Big]  \hspace{0.1cm}+  \nonumber\\
& (1-p_S) \frac{\tr[\mathcal{M}_{\omega}({\pmb{\sigma}}^{US})]}{\tr[\mathcal{M}_{\omega}({\pmb{\sigma}})]}  p(a^{\prime}|x^{\prime})\tr\Big[\Pi_{b|y} [\mathcal{D}_{\omega}({\pmb{\sigma}}^{US})]_{a'|x'} \Big] \nonumber\\
&\forall a^{\prime}, x^{\prime}, b, y.
\end{align} 
Hence, from Eq. (\ref{eq7}), we get the following decomposition for the box $P[\mathcal{D}_{\omega}({\pmb{\sigma}})]$:
 \begin{align}
P[\mathcal{D}_{\omega}({\pmb{\sigma}})] &= p_S \dfrac{\tr[\mathcal{M}_{\omega}({\pmb{\sigma}}^S)]}{\tr[\mathcal{M}_{\omega}({\pmb{\sigma}})]} P[\mathcal{D}_{\omega}({\pmb{\sigma}}^S)] \nonumber \\
&+ (1-p_S)  \dfrac{\tr[\mathcal{M}_{\omega}({\pmb{\sigma}}^{US})]}{\tr[\mathcal{M}_{\omega}({\pmb{\sigma}})]}  P[\mathcal{D}_{\omega}({\pmb{\sigma}}^{US})]. 
\label{aft}
\end{align}
Note that the assemblage $\left\{p(a'|x')[\mathcal{D}_{\omega}({\pmb{\sigma}}^{US})]_{a'|x'}\right\}_{a',x'} \in \Sigma^{US}$
since the assemblage ${\pmb{\sigma}}^{US}$ is unsteerable  \cite{GA15}.
This implies that the box $P[\mathcal{D}_{\omega}({\pmb{\sigma}}^{US})]$ in the decomposition (\ref{aft})
is an unsteerable box. There are now two cases which have to be checked to verify Eq. (\ref{ADLOCC}).
(i) Suppose the assemblage $\left\{p(a'|x')[\mathcal{D}_{\omega}({\pmb{\sigma}}^{S})]_{a'|x'}\right\}_{a',x'}$ is unsteerable.  Then from Eq. (\ref{aft}) it is clear that the box $P[\mathcal{D}_{\omega}({\pmb{\sigma}})]$ is a convex mixture of two unsteerable boxes and, hence, unsteerable. Therefore,
in this case, the following inequality trivially holds:
\begin{align}
&\sum_{\omega} \mathcal{T}(\omega) C_{steer}(P[\mathcal{D}_{\omega}({\pmb{\sigma}})]) \nonumber \\
& = 0  \leq C_{steer}(P[{\pmb{\sigma}}]) \hspace{0.5cm} \forall {\pmb{\sigma}} \in \Sigma^{S}.
\end{align}
(ii) Suppose the assemblage $\left\{p(a'|x')[\mathcal{D}_{\omega}({\pmb{\sigma}}^{S})]_{a'|x'}\right\}_{a',x'}$ is steerable
and  the box $P[\mathcal{D}_{\omega}({\pmb{\sigma}}^S)]$ in  the decomposition (\ref{aft}) is a steerable box.
Then, the decomposition (\ref{aft}) may not be
the optimal decomposition 
(for which the weight of the steerable part being the minimum over all possible decompositions of the box $P[\mathcal{D}_{\omega}({\pmb{\sigma}})]$).
Hence, one has to minimize the weight of the steerable part $p_S \frac{\tr[\mathcal{M}_{\omega}({\pmb{\sigma}}^S)]}{\tr[\mathcal{M}_{\omega}({\pmb{\sigma}})]}$ in Eq. (\ref{aft})
over all possible decompositions of the box $P[\mathcal{D}_{\omega}({\pmb{\sigma}})]$ to obtain the  steering cost $C_{steer}(P[\mathcal{D}_{\omega}({\pmb{\sigma}})])$ of the box. Therefore, we have  
\begin{align}
&C_{steer}(P[\mathcal{D}_{\omega}({\pmb{\sigma}})]) \nonumber\\
&\leq p_S \frac{\tr[\mathcal{M}_{\omega}({\pmb{\sigma}}^S)]}{\tr[\mathcal{M}_{\omega}({\pmb{\sigma}})]} \nonumber \\
&= C_{steer}(P[{\pmb{\sigma}}])  \frac{\tr[\mathcal{M}_{\omega}({\pmb{\sigma}}^S)]}{\tr[\mathcal{M}_{\omega}({\pmb{\sigma}})]}. \label{min1}
\end{align}
The last equality holds as we have assumed that the decomposition (\ref{bef}) denotes the optimal decomposition of the box $P[{\pmb{\sigma}}]$, i.e., $p_{S}=C_{steer}(P[{\pmb{\sigma}}])$. As for all $\omega$, $\mathcal{T}(\omega) = \tr[\mathcal{M}_{\omega} ({\pmb{\sigma}})] \geq 0$ and  $\sum_{\omega} \mathcal{T}(\omega) \leq 1$,  from Eq. (\ref{min1}) we get for deterministic 1W-LOCCs,
\begin{align}
&\sum_{\omega} \mathcal{T}(\omega) C_{steer}(P[\mathcal{D}_{\omega}({\pmb{\sigma}})]) \nonumber \\
&\leq \sum_{\omega} \mathcal{T}(\omega) C_{steer}(P[{\pmb{\sigma}}])  \frac{\tr[\mathcal{M}_{\omega}({\pmb{\sigma}}^S)]}{\tr[\mathcal{M}_{\omega}({\pmb{\sigma}})]} \nonumber\\
&=\sum_{\omega} C_{steer}(P[{\pmb{\sigma}}]) \tr[\mathcal{M}_{\omega}({\pmb{\sigma}}^S)] \nonumber\\
&=C_{steer}(P[{\pmb{\sigma}}]) \sum_{\omega} \tr[\mathcal{M}_{\omega}({\pmb{\sigma}}^S)] \nonumber\\
& \leq C_{steer}(P[{\pmb{\sigma}}]), \hspace{0.5cm} \forall {\pmb{\sigma}} \in \Sigma^{S}.
\end{align}
The last inequality holds, because ${\pmb{\sigma}}^S$ is the set of unnormalized conditional states and 
$\mathcal{M}$ is a deterministic map, i.e.,  $\sum_{\omega} \tr[\mathcal{M}_{\omega}({\pmb{\sigma}}^S)] \leq 1$. This completes the proof for the monotonicity of $C_{steer}(P)$ on average, under 1W-LOCCs for all 
assemblages.
\end{proof}

\item For all convex decompositions of
\be
{\pmb{\sigma}} = \mu {\pmb{\sigma^{\prime}}} + (1-\mu) {\pmb{\sigma^{\prime \prime}}}, \label{conAss}
\ee
in terms of the other two  assemblages ${\pmb{\sigma^{\prime}}}$ and ${\pmb{\sigma^{\prime \prime}}}$
with $0 \leq \mu \leq 1$, 
\begin{align}
C_{steer}(P[{\pmb{\sigma}}]) & \leq \mu C_{steer}(P[{\pmb{\sigma^{\prime}}}]) \nonumber\\
& + (1-\mu) C_{steer}(P[{\pmb{\sigma^{\prime \prime}}}]) \nonumber\\
& \forall {\pmb{\sigma}} \in  \Sigma^{S}.
\end{align}

\begin{proof}
Note that an arbitrary assemblage ${\pmb{\sigma}} :=\{\sigma_{a|x}\}_{a,x} \in \Sigma^S$ satisfies the following relation 
for all possible convex decompositions as in Eq. (\ref{conAss}): 
\begin{align}
\label{d2}
\tr[\Pi_{b|y} \sigma_{a|x}] =& \mu \tr[\Pi_{b|y} \sigma^{\prime}_{a|x}] + (1-\mu) \tr[\Pi_{b|y} \sigma^{\prime\prime}_{a|x}] \nonumber\\
& \forall a,x,b,y, 
\end{align}
which implies that 
the box $P[{\pmb{\sigma}}]$ arising from the assemblage ${\pmb{\sigma}}$
has the following decomposition: 
\begin{align}
P[{\pmb{\sigma}}] = \mu P[{\pmb{\sigma^{\prime}}}] + (1-\mu) P[{\pmb{\sigma^{\prime \prime}}}],
\label{dec1}
\end{align}
where the box $P[{\pmb{\sigma^{\prime}}}]$  arises from the assemblage ${\pmb{\sigma^{\prime}}} :=\{\sigma^{\prime}_{a|x}\}_{a,x}$ and the box $P[{\pmb{\sigma^{\prime \prime}}}]$ arises from the assemblage ${\pmb{\sigma^{\prime\prime}}}:=\{\sigma^{\prime \prime}_{a|x}\}_{a,x}$. We write the optimal decompositions (with weight of the steerable part being the minimum over the all possible decompositions) for the two boxes in the above decomposition (\ref{dec1}) as follows: 
\begin{align}
P[{\pmb{\sigma^{\prime}}}] &:= C_{steer}(P[{\pmb{\sigma^{\prime}}}]) P_S^{1} + (1-C_{steer}(P[{\pmb{\sigma^{\prime}}}])) P_{US}^{1},
\end{align}
where  $P_S^{1}$ and $P_{US}^{1}$ are steerable and unsteerable boxes, respectively, and 
$0\le C_{steer}(P[{\pmb{\sigma^{\prime}}}])\le1$ is the steering cost of the box  $P[{\pmb{\sigma^{\prime}}}]$,
and
\begin{align}
P[{\pmb{\sigma^{\prime \prime}}}] &:= C_{steer}(P[{\pmb{\sigma^{\prime \prime}}}]) P_S^{2} + (1-C_{steer}(P[{\pmb{\sigma^{\prime \prime}}}])) P_{US}^{2},
\end{align} 
where  $P_S^{2}$ and $P_{US}^{2}$ are steerable and unsteerable boxes, respectively, and
$0\le C_{steer}(P[{\pmb{\sigma^{\prime \prime}}}])\le1$ is the steering cost of the box $P[{\pmb{\sigma^{\prime \prime}}}]$.
Decomposing the boxes in the decomposition (\ref{dec1}) with the above two optimal decompositions, we obtain 
\begin{align}
P[{\pmb{\sigma}}] &=\mu \Big[C_{steer}(P[{\pmb{\sigma^{\prime}}}]) P_S^{1} + (1-C_{steer}(P[{\pmb{\sigma^{\prime}}}])) P_{US}^{1}\Big]  \nonumber\\
&+ (1-\mu) \Big[C_{steer}(P[{\pmb{\sigma^{\prime \prime}}}]) P_S^{2} + (1-C_{steer}(P[{\pmb{\sigma^{\prime \prime}}}])) P_{US}^{2}\Big] \\
&:= \nu \mathbb{P} +(1-\nu) \mathbb{P}_{US}, \label{con}
\end{align}
with
\be
\nu:= \mu C_{steer}(P[{\pmb{\sigma^{\prime}}}]) + (1-\mu) C_{steer}(P[{\pmb{\sigma^{\prime \prime}}}]), \label{con1}
\ee
which  satisfies $0\leq\nu\leq1$ and  
\begin{align}
\mathbb{P} :=& \frac{1}{\nu}\Big[\mu C_{steer}(P[{\pmb{\sigma^{\prime}}}]) P_S^{1} + (1-\mu) C_{steer}(P[{\pmb{\sigma^{\prime \prime}}}]) P_S^{2}\Big], \label{dec2}
\end{align}
which may be a steerable or an unsteerable box,
and 
\begin{align}
\mathbb{P}_{US} :=& \frac{1}{1-\nu}\Big[\mu (1-C_{steer}(P[{\pmb{\sigma^{\prime}}}])) P_{US}^{1}  \nonumber \\
&+ (1-\mu) (1-C_{steer}(P[{\pmb{\sigma^{\prime \prime}}}])) P_{US}^{2}\Big],
\end{align}
which is an unsteerable box since any convex mixture of two unsteerable boxes is unsteerable.

Suppose the box $\mathbb{P}$ (\ref{dec2}) is unsteerable.  Then from Eq. (\ref{con}) it is clear that the box $P[{\pmb{\sigma}}]$ is a convex mixture of two unsteerable boxes and, hence, unsteerable. Therefore,
in this case the following inequality trivially holds for all possible convex decompositions as in Eq. (\ref{conAss}) of an arbitrary assemblage ${\pmb{\sigma}}\in \Sigma^{S}$:
\begin{align}
C_{steer}(P[{\pmb{\sigma}}])=0 & \leq  \nu = \mu C_{steer}(P[{\pmb{\sigma^{\prime}}}]) + (1-\mu) C_{steer}(P[{\pmb{\sigma^{\prime \prime}}}]).
\end{align}
Suppose the box $\mathbb{P}$ (\ref{dec2}) is steerable.
Then the decomposition (\ref{con}) is not the optimal one if the weights of both the boxes $P^1_S$ and $P^2_S$ are nonzero (since
the box $\mathbb{P}$ is not an extremal box in this case, because an extremal steerable box in the set $\mathcal{N}_{\Sigma^{S}}$ cannot be decomposed as a convex mixture of the other boxes in the set $\mathcal{N}_{\Sigma^{S}}$).
Even if the weight of the box $P^1_S$ or that of the box $P^2_S$ is zero, the decomposition (\ref{con}) may  not be the optimal one. Hence, to obtain the steering cost $C_{steer}(P[{\pmb{\sigma}}])$ of the box $P[{\pmb{\sigma}}]$,
one has to minimize the weight of the steerable part $\nu$ over all such possible decompositions of the box $P[{\pmb{\sigma}}]$. So we have, 
\be
C_{steer}(P[{\pmb{\sigma}}]) \leq \nu. \label{con2}
\ee
From Eq. (\ref{con2}) together with Eq. (\ref{con1}), we can conclude that for all possible convex decompositions as in Eq. (\ref{conAss}) of an arbitrary assemblage ${\pmb{\sigma}} \in \Sigma^{S}$, 
\begin{align}
C_{steer}(P[{\pmb{\sigma}}]) & \leq \mu C_{steer}(P[{\pmb{\sigma^{\prime}}}]) + (1-\mu) C_{steer}(P[{\pmb{\sigma^{\prime \prime}}}]).
\end{align}
\end{proof}
\end{enumerate}
Since the steering cost $C_{steer}(P)$ satisfies the above two properties, it is a convex steering monotone.

In what follows, we will characterize steerability of two families of correlations which are called white-noise 
  BB84 family and colored-noise BB84 family in the context of the following steering scenario:
 \textit{ Alice performs two black-box dichotomic measurements on her part of an unknown 
  $d\times 2$ quantum state shared with Bob which produce the assemblage $\{\sigma_{a|x}\}_{a,x}$
  on Bob's side. On this assemblage, Bob performs projective qubit measurements $\{\Pi_{b|y}\}_{b,y}$ 
  corresponding to any two mutually unbiased bases (MUBs), 
  i.e. $B_0 \equiv \lbrace |f_i\rangle \rbrace_{i=1}^2$ and $B_1 \equiv \lbrace |g_j\rangle \rbrace_{j=1}^2$ such that, $|\langle f_i | g_j \rangle|^2 = \frac{1}{2} \, \forall i,j$  (here, $ \lbrace |f_i\rangle \rbrace_{i=1}^2$ and $\lbrace |g_j\rangle \rbrace_{j=1}^2$ are two sets of orthonormal basis).}
 In this scenario, the necessary and sufficient condition for quantum steering from Alice to Bob is given by \cite{CFF+15},
\begin{align}
&\sqrt{\langle (A_0 + A_1) B_0 \rangle^2 + \langle (A_0 + A_1) B_1 \rangle^2 } \nonumber \\
&+\sqrt{\langle (A_0 - A_1) B_0 \rangle^2 + \langle (A_0 - A_1) B_1 \rangle^2 } \leq 2. \label{chshst}
\end{align}
This inequality is called the analogous CHSH inequality for quantum steering.

  The white-noise BB84 and colored-noise BB84 families belong to the local polytope of 
  the two-binary-inputs and two-binary-outputs Bell scenario. In order to 
  find out the existence of a LHV-LHS model for the given local correlation,
  we will consider a classical
  simulation model by using shared classical randomness, i.e., 
  a local hidden variable model of finite dimension \cite{DW15}. Suppose a local box $P_L(ab|xy)$ := $\{p_{L}(ab|xy)\}_{a,x,b,y}$ admits 
  the following decomposition:
  \be
   p_L(ab|xy)=\sum^{d_\lambda-1}_{\lambda=0} p(\lambda) p(a|x,\lambda)p(b|y,\lambda) \hspace{0.3cm} \forall a,x,b,y. \label{srdim}
  \ee
  Then it defines a classical simulation model by using shared randomness of dimension
  $d_\lambda$. 
  
  In Ref. \cite{DW15}, the upper bound on the minimum dimension 
  of shared randomness required to simulate a local $n$-partite correlation is derived
  (see Proposition $5$ in Ref. \cite{DW15}). 
  For the bipartite Bell scenario with two-binary inputs and two-binary outputs, 
  shared randomness  of dimension $d_\lambda \le 4$ is sufficient to simulate any local box. 
  
  Our method to check the existence of a LHV-LHS model 
  for the local correlations in terms of the extremal boxes of the given steering scenario goes as follows. We first decompose the given local 
  correlation in the form (\ref{srdim}) where $p(a|x,\lambda)$ are different deterministic distributions and  
  $p(b|y,\lambda)$ may be nondeterministic
  in order to minimize the dimension of shared randomness.
  Then,  we try to check whether each Bob's distribution in this decomposition has a quantum realization in the context of the given steering scenario.
\subsection{White noise BB84 family}

Consider the  family of correlations defined as
\be
P_{BB84}(ab|xy)=\frac{1+(-1)^{a\oplus b \oplus {x \cdot y}}\delta_{x,y}V}{4}, \label{bb84fam}
\ee
where $0<V\le1$. For $V=1$, the above family of correlations corresponds to the BB84 correlation 
\footnote{The BB84 correlation satisfies $p(a=b|xy)=1$ for $x=y$ and $p(a=b|xy)=1/2$ for $x \ne y$, here $p(a=b|xy)=p(00|xy)+p(11|xy)$ 
\cite{AGM06}.} upto LRO.  
For this reason, we refer to the family of correlations given in Eq. (\ref{bb84fam}) as white noise BB84 family.
The white noise BB84 family  is 
local as it does not violate a Bell-CHSH inequality (\ref{BCHSH}).
The white noise BB84 family can be obtained from
the two-qubit Werner state, 
\be
\rho_W=V\ketbra{\Psi^-}{\Psi^-}+\frac{1-V}{4}\openone, \label{werst}
\ee
where $\ket{\Psi^-}=(\ket{01}-\ket{10})/\sqrt{2}$,
with the projectors $\Pi_{a|x=0}=\frac{1}{2}(\openone-a\sigma_z)$, $\Pi_{a|x=1}=\frac{1}{2}(\openone+a\sigma_x)$,
$\Pi_{b|y=0}=\frac{1}{2}(\openone+b\sigma_z)$ and $\Pi_{b|y=1}=\frac{1}{2}(\openone+b\sigma_x)$.
The Werner state is entangled iff $V>1/3$ \cite{Wer89}.\\

 The white noise BB84 family violates the analogous CHSH inequality for quantum steering given by Eq.(\ref{chshst}) for $V > \frac{1}{\sqrt{2}}$. Hence, the white noise BB84 family cannot be decomposed as a convex mixture of the extremal points of the unsteerable set as in Eq. (\ref{DLHS}) iff $V>1/\sqrt{2}$ in the given steering scenario, i. e., where Alice performs two black-box dichotomic measurements and  Bob performs projective qubit measurements corresponding to any two mutually unbiased bases (MUBs). In the following we will demonstrate our procedure to find out in which range the white noise BB84 family can be written as a convex mixture of the extremal points of the unsteerable set as in Eq. (\ref{DLHS}) in the given steering scenario.\\

In the context of nonsignaling polytope, the BB84 family can be decomposed as follows:
\be
P_{BB84}=V\left(\frac{P^{000}_{PR}+P^{110}_{PR}}{2}\right)+(1-V)P_N, \label{PRbb84}
\ee
where $P_N$ is the maximally mixed box, i.e., $P_N (ab|xy) = \frac{1}{4}$, $\forall a,b,x,y$.
Let us rewrite the above decomposition  as follows:
\ba
&&P_{BB84} \nonumber \\
&=& V  \bigg(\frac{1}{2} P_{PR}^{000}
 + \frac{1}{2} P_N \bigg) + V  \bigg(\frac{1}{2} P_{PR}^{110} + \frac{1}{2} P_N \bigg) + (1-2V) P_N. \nonumber \\
 \ea
By writing the each box in the above decomposition in terms of the local deterministic boxes,
we obtain the following decomposition which defines a classical simulation protocol 
by using shared randomness of dimension $4$:
 \ba
&& P_{BB84}(ab|xy)\nonumber \\
&=& \frac{V}{8}\sum_{\alpha\beta\gamma}P^{\alpha\beta\gamma(\alpha\gamma\oplus\beta)}_D(ab|xy)+\frac{V}{8}\sum_{\alpha\beta\gamma}P^{\alpha \beta \gamma(\bar{\alpha}\bar{\gamma}\oplus\beta)}_D(ab|xy) \nonumber \\
&+&(1-2V) P_N \\
&=&\frac{1}{4}\sum_{\lambda=1}^{4}P_\lambda(a|x)P_\lambda(b|y), \label{BB84dim4}
\ea
where  $\bar{\alpha} = \alpha \oplus 1$; $\bar{\gamma} = \gamma \oplus 1$; $P_\lambda(a|x)$:=$\{p(a|x,\lambda)\}_{a,x}$ is the set of conditional probability distributions $p(a|x,\lambda)$ for all possible $a$, $x$; and $P_\lambda(b|y)$:=$\{p(b|y,\lambda)\}_{b,y}$ is the set of conditional probability distributions $p(b|y, \lambda)$ for all possible $b$, $y$.
In the LHV model given in Eq. (\ref{BB84dim4}),  one of the parties  (here, Alice) uses deterministic  
strategies given by:
\begin{align}
&P_1(a|x)=P^{00}_D, P_2(a|x)=P^{01}_D, \nonumber \\
&P_3(a|x)=P^{10}_D, P_4(a|x)=P^{11}_D,
\end{align} 
while the other (here, Bob) uses nondeterministic strategies given by:

\begin{align}
P_1(b|y)&=  V  P_D^{10}  +(1-V) \bigg( \frac{P_D^{00} + P_D^{01} + P_D^{10} + P_D^{11}}{4} \bigg), \nonumber \\
P_2(b|y)&=  V  P_D^{11}  +(1-V) \bigg( \frac{P_D^{00} + P_D^{01} + P_D^{10} + P_D^{11}}{4} \bigg) , \nonumber \\
P_3(b|y)&= V  P_D^{00}  +(1-V) \bigg( \frac{P_D^{00} + P_D^{01} + P_D^{10} + P_D^{11}}{4} \bigg) , \nonumber \\
P_4(b|y)&= V  P_D^{01}  +(1-V) \bigg( \frac{P_D^{00} + P_D^{01} + P_D^{10} + P_D^{11}}{4} \bigg).
\label{ndb1}
\end{align}

Let us now try to find in which range the BB84 family  has a decomposition  
in terms of the extremal points of the unsteerable set
as in Eq. (\ref{DLHS}) from the decomposition given in Eq. (\ref{BB84dim4}).
For this purpose, we try to check in which range each nondeterministic strategy on Bob's 
side in Eq. (\ref{ndb1}) can arise from a pure qubit state  in the given steering scenario, i.e., for the measurements $B_0$
and $B_1$ in two mutually unbiased bases (MUB). With this aim, we note that
each of Bob's nondeterministic strategies can be written in the form,
$P_\lambda(b|y)=\langle \psi _\lambda | {\bf{\Pi}}_B | \psi_\lambda \rangle$  (${\pmb{\Pi}}_B$ := $\{\Pi_{b|y}\}_{b,y}$ corresponds to the set projective  measurements at Bob's side in any two mutually unbiased bases  in Hilbert space $C^2$:  $B_0 \equiv \lbrace |f_i\rangle \rbrace_{i=1}^2$ and $B_1 \equiv \lbrace |g_j\rangle \rbrace_{j=1}^2$ such that, $|\langle f_i | g_j \rangle|^2 = \frac{1}{2} \, \forall i,j$, where $ \lbrace |f_i\rangle \rbrace_{i=1}^2$ and $\lbrace |g_j\rangle \rbrace_{j=1}^2$ are two sets of orthonormal basis),
with the following pure states:
\be
|\psi_1 \rangle = \sqrt{\frac{1+V}{2}} |f_1 \rangle + e^{i \phi_1} \sqrt{\frac{1-V}{2}} |f_2 \rangle,
\ee
where $\cos \phi_1 = - \frac{V}{\sqrt{1+V}\sqrt{1-V}}$,
\be
|\psi_2 \rangle = \sqrt{\frac{1-V}{2}} |f_1 \rangle + e^{i \phi_2} \sqrt{\frac{1+V}{2}} |f_2 \rangle,
\ee
where $\cos \phi_2 =  \frac{V}{\sqrt{1+V}\sqrt{1-V}}$,
\be
|\psi_3 \rangle = \sqrt{\frac{1+V}{2}} |f_1 \rangle + e^{i \phi_3} \sqrt{\frac{1-V}{2}} |f_2 \rangle,
\ee
where $\cos \phi_3 =  \frac{V}{\sqrt{1+V}\sqrt{1-V}}$, and 
\be
|\psi_4 \rangle = \sqrt{\frac{1-V}{2}} |f_1 \rangle + e^{i \phi_4} \sqrt{\frac{1+V}{2}} |f_2 \rangle,
\ee
where $\cos \phi_4 = - \frac{V}{\sqrt{1+V}\sqrt{1-V}}$. For any $|\psi_\lambda \rangle$ given above,
$|\cos \phi_\lambda|\le 1$ iff $V\le 1/\sqrt{2}$.  Note that, in the given steering scenario, the above states are the \textit{only} pure states which give rise to the nondeterministic probability distributions on Bob's side in Eq. (\ref{BB84dim4}). Therefore, we can conclude that the decomposition (\ref{BB84dim4}) represents convex mixture of the extremal points of the unsteerable set as in Eq. (\ref{DLHS}) in the given steering scenario iff $V\le \frac{1}{\sqrt{2}}$.

\begin{thm}
 The steering cost of the white noise BB84 family is given by $C_{steer}(P_{BB84})=\max\{0,\frac{\sqrt{2}V-1}{\sqrt{2}-1}\}$  in the given steering scenario.
\end{thm}
\begin{proof}
Note that for $V\ge1/\sqrt{2}$, the BB84 family can be decomposed  as follows:
\be
P_{BB84}=\frac{\sqrt{2}V-1}{\sqrt{2}-1}P^{Ext}_{S}+\frac{\sqrt{2}(1-V)}{\sqrt{2}-1}P_{US}, \label{SWWer}
\ee
where 
\be \label{exsteerb}
P^{Ext}_{S}=\frac{1+(-1)^{a\oplus b\oplus x \cdot y}\delta_{x,y}}{4} 
\ee
is an extremal steerable box 
as it violates the steering inequality (\ref{chshst})
maximally, and $P_{US}$ is an unsteerable box which has a decomposition as in Eq. (\ref{BB84dim4}) with $V=1/\sqrt{2}$
in terms of the extremal boxes of the given steering scenario. 
We see that the weight $\frac{\sqrt{2}V-1}{\sqrt{2}-1}$ in the decomposition (\ref{SWWer}) is 
nonzero iff $P_{BB84}$ detects steerability. 
 Therefore, the decomposition given in Eq. (\ref{SWWer}) is the optimal decomposition for the BB84 family
for $V\ge1/\sqrt{2}$, because it  is a convex mixture of the 
\textit{extremal} steerable box (in the given steering scenario) and the unsteerable box
with the weight of the steerable part going to zero iff the box is unsteerable.

\end{proof}

We will now verify that, $C_{steer}(P_{BB84})=\max\{0,\frac{\sqrt{2}V-1}{\sqrt{2}-1}\}$ 
is a convex roof measure. From Eq. (\ref{as}), we know that the assemblage ${\pmb{\sigma}}\vert_{\rho_{W}}$
arising from the state $\rho_W$ given in Eq. (\ref{werst}) can be decomposed as follows:

\ba
{\pmb{\sigma}}\vert_{\rho_{W}}&=&\tr_A({\bf{M}}_A \otimes \openone \rho_{W})\nonumber  \\
&=& V \, 
\tr_A({\bf{M}}_A\otimes \openone \ketbra{\Psi^-}{\Psi^-}) \nonumber \\
&+& (1-V) \, \tr_A({\bf{M}}_A \otimes \openone \frac{\openone}{4}) \nonumber\\
&=& V {\pmb{\sigma}}\vert_{\ket{\Psi^-}} + (1-V) {\pmb{\sigma}}\vert_{\frac{\openone}{4}}. 
\ea
Here, ${\pmb{\sigma}}\vert_{\ket{\Psi^-}}$ is the  assemblage arising from the state $|\Psi^-\rangle$, and ${\pmb{\sigma}} \vert_{\frac{\openone}{4}}$ is the assemblage arising from the state $\frac{\openone}{4}$. We now see that for any $V\in[0,1]$, the following relation is satisfied: 
\ba
C_{steer}\left(P[{\pmb{\sigma}}\vert_{\rho_{W}}]\right)&\le& V C_{steer}\left( P[{\pmb{\sigma}}\vert_{\ket{\Psi^-}}]\right)  \nonumber \\
&+& (1-V) C_{steer}\left(P[{\pmb{\sigma}}\vert_{\frac{\openone}{4}}]\right), \label{crmbb84}
\ea
for the measurements that generate the BB84 family. Here,  
$C_{steer}\left(P[{\pmb{\sigma}}\vert_{\rho_{W}}]\right)=C_{steer}(P_{BB84})=\max\{0,\frac{\sqrt{2}V-1}{\sqrt{2}-1}\}$,
$C_{steer}\left(P[{\pmb{\sigma}}\vert_{\ket{\Psi^-}}]\right) =1$ (since $P[{\pmb{\sigma}}\vert_{\ket{\Psi^-}}]$  violates the steering inequality (\ref{chshst}) maximally for the aforementioned measurement settings) and  
$C_{steer}\left(P[{\pmb{\sigma}}\vert_{\frac{\openone}{4}}]\right)=0$
since $\frac{\openone}{4}$ does not have steerability. 
In another way, we can conclude that, if $P[{\pmb{\sigma}}]$ and $P[{\pmb{\sigma^{\prime}}}]$ 
are two boxes belonging to the given 
steering scenario and obeying the following relation: 
\be
P[{\pmb{\sigma}}] = \eta P[{\pmb{\sigma^{\prime}}}] + (1-\eta) P_{US} ,
\ee
with $0 \leq \eta \leq 1$ and $P_{US}$ being an unsteerable box, 
then $P[{\pmb{\sigma^{\prime}}}]$ is more steerable than $P[{\pmb{\sigma}}]$ or  equally steerable to $P[{\pmb{\sigma}}]$, analogous to the case of Bell non-locality as demonstrated in Ref. \cite{Vicente}.

\subsection{Colored noise BB84 family}

Let us now consider the colored-noise BB84 family defined as 
\begin{align}
&P^{col}_{BB84} (ab|xy) = \nonumber \\
&\frac{1+(-1)^{a\oplus b \oplus {x \cdot y}}[\delta_{x,y}V+(1-V)/2]+
(-1)^{a\oplus b \oplus x \oplus y}(1-V)/2}{4}, \label{colBB84}
\end{align}
where $0<V\le1$.
Note that for $V=1$, the above family of correlations corresponds to the BB84 correlation  \cite{AGM06} upto LRO.
The colored-noise BB84 family can be obtained from the
colored-noise two-qubit maximally entangled state, 
\be
\label{colst}
\rho_{col}=V\ketbra{\Psi^-}{\Psi^-}+(1-V)\openone_{col},
\ee
where the color noise $\openone_{col}=(\ket{01}\bra{01}+\ket{10}\bra{10})/2$,
for  suitable projective measurements. The colored-noise BB84 family is
local as it does not violate a Bell-CHSH inequality.\\

 The colored-noise BB84 family violates the analogous CHSH inequality for quantum steering (\ref{chshst}) for $V > 0$. Hence, the colored-noise BB84 family cannot be decomposed as a convex mixture of the extremal points of the unsteerable set as in Eq. (\ref{DLHS}) iff $V>0$ in the given steering scenario. In the following, adopting our procedure, we will find out in which range the colored-noise BB84 family can be written as a convex mixture of the extremal points of the unsteerable set as in Eq. (\ref{DLHS}) in the given steering scenario.\\

In the context of nonsignaling polytope, the colored-noise BB84 family can be decomposed as follows:
\be
P^{col}_{BB84}=V\left(\frac{P^{000}_{PR}+P^{110}_{PR}}{2}\right)+(1-V)P_{US}. \label{PRcolbb84}
\ee
Here, 
\be
P_{US}:=\frac{P^{000}_{PR}+P^{110}_{PR}+P^{010}_{PR}+P^{100}_{PR}}{4}, \nonumber
\ee
which  belongs to the unsteerable set of the steering scenario that we have considered.
There are many possible decompositions for the box $P_{US}$ in terms of local deterministic boxes. 
But all of them do not lead to convex mixtures of extremal boxes of the unsteerable set 
in the given steering scenario for any two projective measurements  ${\pmb{\Pi}}_{B}$:= $\{\Pi_{b|y}\}_{b,y}$ in any two mutually unbiased bases (at Bob's side)  in Hilbert space $C^2$:  $B_0 \equiv \lbrace |f_i\rangle \rbrace_{i=1}^2$ and $B_1 \equiv \lbrace |g_j\rangle \rbrace_{j=1}^2$ such that, $|\langle f_i | g_j \rangle|^2 = \frac{1}{2} \, \forall i,j$ (Here $ \lbrace |f_i\rangle \rbrace_{i=1}^2$ and $\lbrace |g_j\rangle \rbrace_{j=1}^2$ are two sets of orthonormal basis). To obtain a such a convex mixture, 
we consider the following decomposition for the box $P_{US}$:
\begin{align}
P_{US}&=\frac{1}{4}P^{00}_D \frac{P_D^{00} + P_D^{10}}{2}+\frac{1}{4}P^{01}_D \frac{P_D^{01} + P_D^{11}}{2} \nonumber \\
& +\frac{1}{4}P^{10}_D \frac{P_D^{00} + P_D^{10}}{2} +\frac{1}{4}P^{11}_D \frac{P_D^{01} + P_D^{11}}{2} \label{uscol0}  \\
&:=\frac{1}{4}P^{00}_D\braket{f_1|{\pmb{\Pi}}_{B}|f_1}+\frac{1}{4}P^{01}_D\braket{f_2|{\pmb{\Pi}}_{B}|f_2} \nonumber \\
& +\frac{1}{4}P^{10}_D\braket{f_1|{\pmb{\Pi}}_B|f_1} +\frac{1}{4}P^{11}_D\braket{f_2|{\pmb{\Pi}}_{B}|f_2}, \label{uscol}
\end{align}
which is a LHV-LHS model in terms of the extremal boxes of the unsteerable set.

By decomposing the first box in the decomposition (\ref{PRcolbb84}) in terms of local deterministic boxes  
and using the decomposition (\ref{uscol0}) for the second box in the decomposition (\ref{PRcolbb84}), we obtain the following LHV model for the colored-noise BB84 box by using shared randomness of dimension $4$:
 \be
P^{col}_{BB84}(ab|xy)
=\frac{1}{4}\sum_{\lambda=1}^{4}P_\lambda(a|x)P_\lambda(b|y), \label{colBB84dim4}
\ee
where  one of the parties  (here, Alice) uses deterministic  
strategies given by:
\begin{align}
&P_1(a|x)=P^{00}_D, P_2(a|x)=P^{01}_D, \nonumber \\
&P_3(a|x)=P^{10}_D, P_4(a|x)=P^{11}_D,
\end{align} 
while the other (here, Bob) uses nondeterministic strategies given by:
\begin{align}
P_1(b|y)&=  V P_D^{10} + (1-V) \frac{P_D^{00} + P_D^{10}}{2}, \nonumber \\
P_2(b|y)&=  V P_D^{11} + (1-V) \frac{P_D^{01} + P_D^{11}}{2}, \nonumber \\
P_3(b|y)&=  V P_D^{00} + (1-V) \frac{P_D^{00} + P_D^{10}}{2}, \nonumber \\
P_4(b|y)&=  V P_D^{01} + (1-V) \frac{P_D^{01} + P_D^{11}}{2}.
\label{ndb2}
\end{align}
Let us now try to find in which range the colored-noise BB84 family  has a decomposition  
in terms of the extremal points of the unsteerable set
as in Eq. (\ref{DLHS}) from the decomposition given in Eq. (\ref{colBB84dim4}).
For this purpose, we try to check in which range each nondeterministic strategy on Bob's 
side in Eq. (\ref{ndb2}) can arise from a pure qubit state  in the given steering scenario, i.e., for the measurements $B_0$
and $B_1$ in two mutually unbiased bases (MUB). With this aim, we note that
each of Bob's nondeterministic strategies can be written in the form,
 $P_\lambda(b|y)=\langle \psi^{'} _\lambda | {\pmb{\Pi}}_B | \psi^{'}_\lambda \rangle$ (${\pmb{\Pi}}_B$ := $\{\Pi_{b|y}\}_{b,y}$ corresponds to the set projective qubit measurements at Bob's side in any two mutually unbiased bases: $B_0 \equiv \lbrace |f_i\rangle \rbrace_{i=1}^2$ and $B_1 \equiv \lbrace |g_j\rangle \rbrace_{j=1}^2$ as defined earlier),
with the following pure states:
\be
|\psi^{'}_1 \rangle = \sqrt{\frac{1-V}{2}} |g_1 \rangle + e^{i \phi^{'}_1} \sqrt{\frac{1+V}{2}} |g_2 \rangle,
\ee
where $\cos \phi^{'}_1 =  \frac{1}{\sqrt{1+V}\sqrt{1-V}}$,
\be
|\psi^{'}_2 \rangle = \sqrt{\frac{1+V}{2}} |g_1 \rangle + e^{i \phi^{'}_2} \sqrt{\frac{1-V}{2}} |g_2 \rangle,
\ee
where $\cos \phi^{'}_2 =  \frac{-1}{\sqrt{1+V}\sqrt{1-V}}$,
\be
|\psi^{'}_3 \rangle = \sqrt{\frac{1+V}{2}} |g_1 \rangle + e^{i \phi^{'}_3} \sqrt{\frac{1-V}{2}} |g_2 \rangle,
\ee
where $\cos \phi^{'}_3 =  \frac{1}{\sqrt{1+V}\sqrt{1-V}}$, and 
\be
|\psi^{'}_4 \rangle = \sqrt{\frac{1-V}{2}} |g_1 \rangle + e^{i \phi^{'}_4} \sqrt{\frac{1+V}{2}} |g_2 \rangle,
\ee
where $\cos \phi^{'}_4 =  \frac{-1}{\sqrt{1+V}\sqrt{1-V}}$. For any $|\psi_\lambda \rangle$ given above,
$|\cos \phi^{'}_\lambda|\le 1$ iff $V=0$.  Note that, in the given steering scenario, the above states are the \textit{only} pure states which give rise to the nondeterministic probability distributions on Bob's side in Eq. (\ref{colBB84dim4}). Therefore, we can conclude that the decomposition (\ref{colBB84dim4}) represents  convex mixture of the extremal points of the unsteerable set as in Eq. (\ref{DLHS}) in the given steering scenario iff $V=0$.

\begin{thm}
The steering cost of the colored-noise BB84 family is given by $C_{steer}(P^{col}_{BB84})=V$  in the given steering scenario.
\end{thm}
\begin{proof}
Note that the colored-noise BB84 family can be decomposed as follows:
\be
\label{wnbb}
P^{col}_{BB84}=VP^{Ext}_{S}+(1-V)P_{US},
\ee
where  $P^{Ext}_{S}$ is the extremal steerable box given in Eq. (\ref{exsteerb})
and $P_{US}$ is the unsteerable box given in Eq. (\ref{uscol}).  The decomposition given in Eq. (\ref{wnbb}) is the optimal decomposition for the BB84 family, because it is a convex mixture of the \textit{extremal} steerable box (in the given steering scenario) and the unsteerable box with the weight of steerable part goes to zero 
iff the box is unsteerable.
\end{proof}

We will now verify that, $C_{steer}(P^{col}_{BB84})=V$ 
is a convex roof measure. Note that the assemblage ${\pmb{\sigma}}\vert_{\rho_{col}}$ arising 
from state $\rho_{col}$ given in Eq. (\ref{colst}) can be decomposed as follows:
\ba
{\pmb{\sigma}}\vert_{\rho_{col}}&=&\tr_A({\bf{M}}_A\otimes \openone \rho_{col}) \nonumber \\
&=& V \, 
\tr_A({\bf{M}}_A\otimes \openone \ketbra{\Psi^-}{\Psi^-}) \nonumber \\
&+& (1-V) \, \tr_A({\bf{M}}_A\otimes \openone \openone_{col}) \nonumber \\
&=& V {\pmb{\sigma}}\vert_{\ketbra{\Psi^-}{\Psi^-}} + (1-V) {\pmb{\sigma}} \vert_{\openone_{col}}.
\ea
Here, ${\pmb{\sigma}}\vert_{\ket{\Psi^-}}$ is the assemblage arising from the state $|\Psi^-\rangle$, and ${\pmb{\sigma}} \vert_{\openone_{col}}$ is the assemblage arising from the state $\openone_{col}$. We now see that for any $V\in[0,1]$, the following relation is satisfied: 
\ba
C_{steer}\left(P[{\pmb{\sigma}}\vert_{\rho_{col}}]\right)&=& V C_{steer}\left( P[{\pmb{\sigma}}\vert_{\ket{\Psi^-}}]\right)  \nonumber \\
&+& (1-V) C_{steer}\left(P[{\pmb{\sigma}}\vert_{\openone_{col}}]\right),
\ea
for the measurements that generate the colored-noise BB84 family. Here,  
$C_{steer}\left(P[{\pmb{\sigma}} \vert_{\rho_{col}}]\right)=C_{steer}(P^{col}_{BB84})=V$, $C_{steer}\left(P[{\pmb{\sigma}} \vert_{\ket{\Psi^-}}]\right) =1$ since $P[{\pmb{\sigma}}\vert_{\ket{\Psi^-}}]$ is an extremal box 
(it violates the steering inequality (\ref{chshst}) maximally for the aforementioned measurement settings)
and  
$C_{steer}\left(P[{\pmb{\sigma}} \vert_{\openone_{col}}]\right)=0$
since $\openone_{col}$ does not have steerability.

\section{Steering cost versus steering weight} \label{scsw}

For any assemblage ${\pmb{\sigma}}=\{\sigma_{a|x}\}_{a,x}$ arising from a given steering scenario, steering weight  \cite{SNC14}
which we denote by $W_{steer}({\pmb{\sigma}})$ is defined as follows.
Consider the following decomposition of the given assemblage ${\pmb{\sigma}}$:
\begin{equation}
\sigma_{a|x}=p_S  \sigma^S_{a \vert x} + (1-p_S) \sigma^{US}_{a \vert x} \hspace{0.2cm} \forall a,x, \label{een1}
\end{equation}
where $\sigma^S_{a \vert x}$ is an element of an   assemblage  ${\pmb{\sigma}}^S$ having steerability and $\sigma^{US}_{a \vert x}$ is an element of an unsteerable assemblage ${\pmb{\sigma}}^{US}$. 
The weight of the steerable part $p_S$ minimized over all possible decompositions of the given assemblage ${\pmb{\sigma}}$
gives the steering weight $W_{steer}({\pmb{\sigma}})$.

\begin{prop}
Let us assume the following optimal decomposition of the given assemblage ${\pmb{\sigma}}=\{\sigma_{a|x}\}_{a,x}$ with the weight of the steerable part being minimized over all possible decompositions of the assemblage, i. e., the weight of the steerable part being equal to the steering weight $W_{steer}({\pmb{\sigma}})$ of the assemblage ${\pmb{\sigma}}$:
 \begin{equation}
\sigma_{a|x}=W_{steer}({\pmb{\sigma}})  \tilde{\sigma}^S_{a \vert x} + (1-W_{steer}({\pmb{\sigma}})) \tilde{\sigma}^{US}_{a \vert x} \hspace{0.2cm} \forall a,x. \label{een}
\end{equation}
and Bob performs a set of projective measurements ${\bf{\Pi}}_B:=\{\Pi_{b|y}\}_{b,y}$ on ${\pmb{\sigma}}$, we have

 \begin{equation}
C_{steer}\left(P[{\pmb{\sigma}}]\right) \le W_{steer}({\pmb{\sigma}}),
\end{equation}
where  $P[{\pmb{\sigma}}]=P(ab|xy)$=$\Big\{ \tr\big[\Pi_{b|y} \sigma_{a|x} \big] \Big\}_{a,x,b,y}$; $\tilde{\sigma}^S_{a \vert x}$ is an element of an   assemblage  ${\pmb{\tilde{\sigma}}}^S$ having steerability and $\tilde{\sigma}^{US}_{a \vert x}$ is an element of an unsteerable assemblage ${\pmb{\tilde{\sigma}}}^{US}$
\end{prop}
\begin{proof}
Suppose Bob performs a set of projective measurements ${\bf{\Pi}}_B:=\{\Pi_{b|y}\}_{b,y}$ on ${\pmb{\sigma}}$
given by the decomposition (\ref{een}). Then, one can write, 
\begin{align}
\label{een2}
\tr\big[\Pi_{b|y} \sigma_{a|x} \big]  &= W_{steer}({\pmb{\sigma}}) \tr\big[\Pi_{b|y} \tilde{\sigma}^S_{a|x} \big] \\
&+ (1-W_{steer}({\pmb{\sigma}})) \tr\big[\Pi_{b|y} \tilde{\sigma}^{US}_{a|x} \big], \quad \forall a,x,b,y. \nonumber
\end{align}
Hence, for the box $P[{\pmb{\sigma}}]$ arising from the assemblage ${\pmb{\sigma}}$, one can write the following decomposition: 
\begin{align}
P[{\pmb{\sigma}}]  &= W_{steer}({\pmb{\sigma}}) P_S[{\pmb{\tilde{\sigma}}}^S] + (1-W_{steer}({\pmb{\sigma}})) P_{US}[{\pmb{\tilde{\sigma}}}^{US}]. \label{een3}
\end{align}
Here, $P_S[{\pmb{\tilde{\sigma}}}^S]$  is a steerable box, produced from the steerable assemblage ${\pmb{\tilde{\sigma}}}^S$ and 
$P_S[{\pmb{\tilde{\sigma}}}^{US}]$ is an unsteerable box, produced from the unsteerable assemblage ${\pmb{\tilde{\sigma}}}^{US}$.

Now in the decomposition (\ref{een3}) the weight $W_{steer}({\pmb{\sigma}})$ of the steerable correlation may not be minimum weight of the steerable correlation over all possible decompositions of the correlation $P[{\pmb{\sigma}}]$. Since the steering cost of the correlation $P[{\pmb{\sigma}}]$ is obtained by minimizing the weight of the steerable correlation over all possible decompositions of the correlation $P[{\pmb{\sigma}}]$, the steering cost $C_{steer}(P[{\pmb{\sigma}}])$ of the correlation $P[{\pmb{\sigma}}]$
satisfies the relationship given by $C_{steer}(P[{\pmb{\sigma}}]) \leq W_{steer}({\pmb{\sigma}})$.
\end{proof}


We will now present two examples demonstrating the above proposition. Suppose Alice and Bob share the two-qubit Werner state $\rho_W(V)$ given by Eq.(\ref{werst}) and Alice performs projective measurements ${\bf{M}}_A:=\{M_{a|x}\}_{a,x}$ in the two bases: $\{ \frac{1}{2}(\openone-\sigma_z)$, $\frac{1}{2}(\openone+\sigma_z) \}$ and $\{ \frac{1}{2}(\openone+\sigma_x)$, $\frac{1}{2}(\openone-\sigma_x)\}$. 
Then the assemblage prepared on Bob's side which we denote by ${\pmb{\sigma}} \vert_{\rho_{W}}$ is steerable iff $V > \frac{1}{\sqrt{2}}$ \cite{CJW+09, CS17}.  For $V \ge \frac{1}{\sqrt{2}}$, this assemblage can be decomposed in the following way,
\begin{align}
\sigma_{a|x}\vert_{\rho_W(V)}& = \frac{\sqrt{2}V-1}{\sqrt{2}-1}\sigma_{a|x}\vert_{\ket{\psi^-}} \nonumber \\
&+\frac{\sqrt{2}(1-V)}{\sqrt{2}-1}\sigma_{a|x}\vert_{\rho_W (V= 1/\sqrt{2})} \hspace{0.2cm} \forall a,x,
\label{swe1}
\end{align}
where $\sigma_{a|x}\vert_{\ket{\psi^-}}$ represents the element of assemblage prepared on Bob's side when Alice performs the aforementioned measurements on the singlet state $\ket{\Psi^-}=(\ket{01}-\ket{10})/\sqrt{2}$, which is an element of steerable assemblage and $\sigma_{a|x}\vert_{\rho_W (V= 1/\sqrt{2})}$ represents the element of assemblage prepared on Bob's side when Alice performs the aforementioned measurements on the shared two-qubit Werner state $\rho_W$ (\ref{werst}) for $V=\frac{1}{\sqrt{2}}$, which is an element of unsteerable assemblage \cite{CS17}. It can be checked that, for all $a$, $x$, each element of the steerable assemblage $\sigma_{a|x}\vert_{\ket{\psi^-}}$ is a pure state after normalization and hence, cannot be written as a convex combination of steerable and unsteerable assemblage. The coefficient of the element of the steerable assemblage in the decomposition (\ref{swe1}), therefore, cannot be reduced further. Moreover, the weight of steerable part goes to zero 
iff the assemblage ${\pmb{\sigma}} \vert_{\rho_{W}}$ is unsteerable. Hence, the decomposition (\ref{swe1}) is the optimal decomposition of the assemblage ${\pmb{\sigma}} \vert_{\rho_{W}}$.  This implies that the steering weight of the two-qubit Werner state $\rho_W$, when Alice performs the aforementioned two measurements, is given by $\max\{0,\frac{\sqrt{2}V-1}{\sqrt{2}-1}\}$.

If Bob performs projective measurements ${\bf{\Pi}}_B:=\{\Pi_{b|y}\}_{b,y}$ in the two mutually unbiased bases: $\{ \frac{1}{2}(\openone+\sigma_z)$, $\frac{1}{2}(\openone-\sigma_z) \}$ and $\{ \frac{1}{2}(\openone+\sigma_x)$, $\frac{1}{2}(\openone-\sigma_x)\}$ on the above assemblage ${\pmb{\sigma}} \vert_{\rho_{W}}$, then the white noise BB84 family is produced. The steering cost of the white noise BB84 family is given by $\max\{0,\frac{\sqrt{2}V-1}{\sqrt{2}-1}\}$. Hence, with these measurements performed by Alice and Bob, the steering cost of the state $\rho_{W}$ 
is equal to the steering weight of the state $\rho_{W}$.

Now, instead of performing the above measurements, if Bob performs projective measurements ${\bf{\Pi}}_B:=\{\Pi_{b|y}\}_{b,y}$ in the two mutually unbiased bases: $\{ \frac{1}{2}(\openone+ cos\frac{\pi}{5} \sigma_x + sin \frac{\pi}{5} \sigma_y)$, $\frac{1}{2}(\openone-cos\frac{\pi}{5} \sigma_x + sin \frac{\pi}{5} \sigma_y) \}$ and $\{ \frac{1}{2}(\openone+\sigma_z)$, $\frac{1}{2}(\openone-\sigma_z)\}$ on the above assemblage ${\pmb{\sigma}} \vert_{\rho_{W}}$, then the produced correlation violates analogous CHSH inequality for quantum steering (\ref{chshst}) for $V > \sqrt{\frac{2}{29}(11-\sqrt{5})}$. Hence, for $0 < V \leq \sqrt{\frac{2}{29}(11-\sqrt{5})}$, the steering cost of the produced correlation is $0$. In the range $\frac{1}{\sqrt{2}} < V \leq \sqrt{\frac{2}{29}(11-\sqrt{5})}$, with these measurements performed by Alice and Bob, the steering cost of the correlation is, therefore, less than the steering weight of the  assemblage from which this correlation has been produced in the given steering scenario.


Experimentally, the determination of  the steering weight for the steering scenario
that we have considered
requires complete tomographic knowledge of the qubit assemblage prepared on the trusted side \cite{JKM+01}. On the 
other hand,
the steering cost proposed by us is determined 
from the observed correlations without having the complete tomographic knowledge of the assemblage prepared.
Thus, the determination of our steering cost is experimentally less demanding than the determination of the
steering weight.

\section{Conclusions} \label{conc}

In this work, we have presented a method to check steerability for the scenario where Alice performs two black-box dichotomic measurements and Bob 
performs  two arbitrary projective qubit measurements in mutually unbiased bases (MUBs). 
This method is based on the decompositions of  
the measurement correlations in the context of the extremal boxes of the  steering scenario.
Our method provides a simple way to check the existence of a LHV-LHS model
for the  measurement correlations. 
Based on this formulation to check steerability, we have proposed a quantifier of steering 
called steering cost.
The determination of our steering cost is experimentally less demanding than the determination of the
steering weight.
We have demonstrated that our steering cost is a convex steering monotone.
We have illustrated our method to check steerability with two families of measurement correlations
and obtained their steering cost.
In Ref. \cite{AGM06}, security of the device-independent quantum key distribution protocol with the nonlocal correlations arising from the 
two-qubit Werner states was studied in the context of extremal nonsignaling boxes.
Similarly, it would be interesting to study security of the one-sided device-independent
quantum key distribution protocol with the measurement correlations that we have considered 
in the context of extremal 
boxes of the steering scenario.

\section*{Acknowledgements}
DD acknowledges the financial support from University
Grants Commission (UGC), Government of India. CJ and ASM  thank  Paul Skrzypczyk for his comments and acknowledge  support through the project SR/S2/LOP-08/2013 
of the DST, Govt.  of India.
The authors acknowledge the anonymous referees for their helpful comments.
S.D. acknowledges financial support through INSPIRE Fellowship from DST, India (Grant No. C/5576/IFD/2015-16).
\bibliography{JM}

\begin{thebibliography}{37}%
\makeatletter
\providecommand \@ifxundefined [1]{%
 \@ifx{#1\undefined}
}%
\providecommand \@ifnum [1]{%
 \ifnum #1\expandafter \@firstoftwo
 \else \expandafter \@secondoftwo
 \fi
}%
\providecommand \@ifx [1]{%
 \ifx #1\expandafter \@firstoftwo
 \else \expandafter \@secondoftwo
 \fi
}%
\providecommand \natexlab [1]{#1}%
\providecommand \enquote  [1]{``#1''}%
\providecommand \bibnamefont  [1]{#1}%
\providecommand \bibfnamefont [1]{#1}%
\providecommand \citenamefont [1]{#1}%
\providecommand \href@noop [0]{\@secondoftwo}%
\providecommand \href [0]{\begingroup \@sanitize@url \@href}%
\providecommand \@href[1]{\@@startlink{#1}\@@href}%
\providecommand \@@href[1]{\endgroup#1\@@endlink}%
\providecommand \@sanitize@url [0]{\catcode `\\12\catcode `\$12\catcode
  `\&12\catcode `\#12\catcode `\^12\catcode `\_12\catcode `\%12\relax}%
\providecommand \@@startlink[1]{}%
\providecommand \@@endlink[0]{}%
\providecommand \url  [0]{\begingroup\@sanitize@url \@url }%
\providecommand \@url [1]{\endgroup\@href {#1}{\urlprefix }}%
\providecommand \urlprefix  [0]{URL }%
\providecommand \Eprint [0]{\href }%
\providecommand \doibase [0]{http://dx.doi.org/}%
\providecommand \selectlanguage [0]{\@gobble}%
\providecommand \bibinfo  [0]{\@secondoftwo}%
\providecommand \bibfield  [0]{\@secondoftwo}%
\providecommand \translation [1]{[#1]}%
\providecommand \BibitemOpen [0]{}%
\providecommand \bibitemStop [0]{}%
\providecommand \bibitemNoStop [0]{.\EOS\space}%
\providecommand \EOS [0]{\spacefactor3000\relax}%
\providecommand \BibitemShut  [1]{\csname bibitem#1\endcsname}%
\let\auto@bib@innerbib\@empty
\bibitem [{\citenamefont {Bell}(1964)}]{Bel64}%
  \BibitemOpen
  \bibfield  {author} {\bibinfo {author} {\bibfnamefont {J.~S.}\ \bibnamefont
  {Bell}},\ }\bibfield  {title} {\enquote {\bibinfo {title} {On the
  einstein-podolsky-rosen paradox.}}\ }\href@noop {} {\bibfield  {journal}
  {\bibinfo  {journal} {Physics}\ }\textbf {\bibinfo {volume} {1}},\ \bibinfo
  {pages} {195} (\bibinfo {year} {1964})}\BibitemShut {NoStop}%
\bibitem [{\citenamefont {Brunner}\ \emph {et~al.}(2014)\citenamefont
  {Brunner}, \citenamefont {Cavalcanti}, \citenamefont {Pironio}, \citenamefont
  {Scarani},\ and\ \citenamefont {Wehner}}]{BCP+14}%
  \BibitemOpen
  \bibfield  {author} {\bibinfo {author} {\bibfnamefont {Nicolas}\ \bibnamefont
  {Brunner}}, \bibinfo {author} {\bibfnamefont {Daniel}\ \bibnamefont
  {Cavalcanti}}, \bibinfo {author} {\bibfnamefont {Stefano}\ \bibnamefont
  {Pironio}}, \bibinfo {author} {\bibfnamefont {Valerio}\ \bibnamefont
  {Scarani}}, \ and\ \bibinfo {author} {\bibfnamefont {Stephanie}\ \bibnamefont
  {Wehner}},\ }\bibfield  {title} {\enquote {\bibinfo {title} {Bell
  nonlocality},}\ }\href {\doibase 10.1103/RevModPhys.86.419} {\bibfield
  {journal} {\bibinfo  {journal} {Rev. Mod. Phys.}\ }\textbf {\bibinfo {volume}
  {86}},\ \bibinfo {pages} {419--478} (\bibinfo {year} {2014})}\BibitemShut
  {NoStop}%
\bibitem [{\citenamefont {Schrodinger}(1935)}]{Schrodinger}%
  \BibitemOpen
  \bibfield  {author} {\bibinfo {author} {\bibfnamefont {E.}~\bibnamefont
  {Schrodinger}},\ }\bibfield  {title} {\enquote {\bibinfo {title} {Discussion
  of probability relations between separated systems},}\ }\href {\doibase
  10.1017/S0305004100013554} {\bibfield  {journal} {\bibinfo  {journal}
  {Mathematical Proceedings of the Cambridge Philosophical Society}\ }\textbf
  {\bibinfo {volume} {31}},\ \bibinfo {pages} {555--563} (\bibinfo {year}
  {1935})}\BibitemShut {NoStop}%
\bibitem [{\citenamefont {Einstein}\ \emph {et~al.}(1935)\citenamefont
  {Einstein}, \citenamefont {Podolsky},\ and\ \citenamefont {Rosen}}]{EPR35}%
  \BibitemOpen
  \bibfield  {author} {\bibinfo {author} {\bibfnamefont {A.}~\bibnamefont
  {Einstein}}, \bibinfo {author} {\bibfnamefont {B.}~\bibnamefont {Podolsky}},
  \ and\ \bibinfo {author} {\bibfnamefont {N.}~\bibnamefont {Rosen}},\
  }\bibfield  {title} {\enquote {\bibinfo {title} {Can quantum-mechanical
  description of physical reality be considered complete?}}\ }\href {\doibase
  10.1103/PhysRev.47.777} {\bibfield  {journal} {\bibinfo  {journal} {Phys.
  Rev.}\ }\textbf {\bibinfo {volume} {47}},\ \bibinfo {pages} {777--780}
  (\bibinfo {year} {1935})}\BibitemShut {NoStop}%
\bibitem [{\citenamefont {Wiseman}\ \emph {et~al.}(2007)\citenamefont
  {Wiseman}, \citenamefont {Jones},\ and\ \citenamefont {Doherty}}]{WJD07}%
  \BibitemOpen
  \bibfield  {author} {\bibinfo {author} {\bibfnamefont {H.~M.}\ \bibnamefont
  {Wiseman}}, \bibinfo {author} {\bibfnamefont {S.~J.}\ \bibnamefont {Jones}},
  \ and\ \bibinfo {author} {\bibfnamefont {A.~C.}\ \bibnamefont {Doherty}},\
  }\bibfield  {title} {\enquote {\bibinfo {title} {Steering, entanglement,
  nonlocality, and the einstein-podolsky-rosen paradox},}\ }\href {\doibase
  10.1103/PhysRevLett.98.140402} {\bibfield  {journal} {\bibinfo  {journal}
  {Phys. Rev. Lett.}\ }\textbf {\bibinfo {volume} {98}},\ \bibinfo {pages}
  {140402} (\bibinfo {year} {2007})}\BibitemShut {NoStop}%
\bibitem [{\citenamefont {Reid}(1989)}]{Rei89}%
  \BibitemOpen
  \bibfield  {author} {\bibinfo {author} {\bibfnamefont {M.~D.}\ \bibnamefont
  {Reid}},\ }\bibfield  {title} {\enquote {\bibinfo {title} {Demonstration of
  the einstein-podolsky-rosen paradox using nondegenerate parametric
  amplification},}\ }\href {\doibase 10.1103/PhysRevA.40.913} {\bibfield
  {journal} {\bibinfo  {journal} {Phys. Rev. A}\ }\textbf {\bibinfo {volume}
  {40}},\ \bibinfo {pages} {913--923} (\bibinfo {year} {1989})}\BibitemShut
  {NoStop}%
\bibitem [{\citenamefont {Walborn}\ \emph {et~al.}(2011)\citenamefont
  {Walborn}, \citenamefont {Salles}, \citenamefont {Gomes}, \citenamefont
  {Toscano},\ and\ \citenamefont {Souto~Ribeiro}}]{WSG+11}%
  \BibitemOpen
  \bibfield  {author} {\bibinfo {author} {\bibfnamefont {S.~P.}\ \bibnamefont
  {Walborn}}, \bibinfo {author} {\bibfnamefont {A.}~\bibnamefont {Salles}},
  \bibinfo {author} {\bibfnamefont {R.~M.}\ \bibnamefont {Gomes}}, \bibinfo
  {author} {\bibfnamefont {F.}~\bibnamefont {Toscano}}, \ and\ \bibinfo
  {author} {\bibfnamefont {P.~H.}\ \bibnamefont {Souto~Ribeiro}},\ }\bibfield
  {title} {\enquote {\bibinfo {title} {Revealing hidden einstein-podolsky-rosen
  nonlocality},}\ }\href {\doibase 10.1103/PhysRevLett.106.130402} {\bibfield
  {journal} {\bibinfo  {journal} {Phys. Rev. Lett.}\ }\textbf {\bibinfo
  {volume} {106}},\ \bibinfo {pages} {130402} (\bibinfo {year}
  {2011})}\BibitemShut {NoStop}%
\bibitem [{\citenamefont {Chowdhury}\ \emph {et~al.}(2014)\citenamefont
  {Chowdhury}, \citenamefont {Pramanik}, \citenamefont {Majumdar},\ and\
  \citenamefont {Agarwal}}]{CPM+14}%
  \BibitemOpen
  \bibfield  {author} {\bibinfo {author} {\bibfnamefont {Priyanka}\
  \bibnamefont {Chowdhury}}, \bibinfo {author} {\bibfnamefont {Tanumoy}\
  \bibnamefont {Pramanik}}, \bibinfo {author} {\bibfnamefont {A.~S.}\
  \bibnamefont {Majumdar}}, \ and\ \bibinfo {author} {\bibfnamefont {G.~S.}\
  \bibnamefont {Agarwal}},\ }\bibfield  {title} {\enquote {\bibinfo {title}
  {Einstein-podolsky-rosen steering using quantum correlations in non-gaussian
  entangled states},}\ }\href {\doibase 10.1103/PhysRevA.89.012104} {\bibfield
  {journal} {\bibinfo  {journal} {Phys. Rev. A}\ }\textbf {\bibinfo {volume}
  {89}},\ \bibinfo {pages} {012104} (\bibinfo {year} {2014})}\BibitemShut
  {NoStop}%
\bibitem [{\citenamefont {Oppenheim}\ and\ \citenamefont
  {Wehner}(2010)}]{OW10}%
  \BibitemOpen
  \bibfield  {author} {\bibinfo {author} {\bibfnamefont {Jonathan}\
  \bibnamefont {Oppenheim}}\ and\ \bibinfo {author} {\bibfnamefont {Stephanie}\
  \bibnamefont {Wehner}},\ }\bibfield  {title} {\enquote {\bibinfo {title} {The
  uncertainty principle determines the nonlocality of quantum mechanics},}\
  }\href {\doibase doi.org/10.1126/science.1192065} {\bibfield  {journal}
  {\bibinfo  {journal} {Science}\ }\textbf {\bibinfo {volume} {330}},\ \bibinfo
  {pages} {1072} (\bibinfo {year} {2010})}\BibitemShut {NoStop}%
\bibitem [{\citenamefont {Pramanik}\ \emph {et~al.}(2014)\citenamefont
  {Pramanik}, \citenamefont {Kaplan},\ and\ \citenamefont {Majumdar}}]{PKM14}%
  \BibitemOpen
  \bibfield  {author} {\bibinfo {author} {\bibfnamefont {Tanumoy}\ \bibnamefont
  {Pramanik}}, \bibinfo {author} {\bibfnamefont {Marc}\ \bibnamefont {Kaplan}},
  \ and\ \bibinfo {author} {\bibfnamefont {A.~S.}\ \bibnamefont {Majumdar}},\
  }\bibfield  {title} {\enquote {\bibinfo {title} {Fine-grained
  einstein-podolsky-rosen\char21{}steering inequalities},}\ }\href {\doibase
  10.1103/PhysRevA.90.050305} {\bibfield  {journal} {\bibinfo  {journal} {Phys.
  Rev. A}\ }\textbf {\bibinfo {volume} {90}},\ \bibinfo {pages} {050305 (R)}
  (\bibinfo {year} {2014})}\BibitemShut {NoStop}%
\bibitem [{\citenamefont {Chowdhury}\ \emph {et~al.}(2015)\citenamefont
  {Chowdhury}, \citenamefont {Pramanik},\ and\ \citenamefont
  {Majumdar}}]{CPM15}%
  \BibitemOpen
  \bibfield  {author} {\bibinfo {author} {\bibfnamefont {Priyanka}\
  \bibnamefont {Chowdhury}}, \bibinfo {author} {\bibfnamefont {Tanumoy}\
  \bibnamefont {Pramanik}}, \ and\ \bibinfo {author} {\bibfnamefont {A.~S.}\
  \bibnamefont {Majumdar}},\ }\bibfield  {title} {\enquote {\bibinfo {title}
  {Stronger steerability criterion for more uncertain continuous-variable
  systems},}\ }\href {\doibase 10.1103/PhysRevA.92.042317} {\bibfield
  {journal} {\bibinfo  {journal} {Phys. Rev. A}\ }\textbf {\bibinfo {volume}
  {92}},\ \bibinfo {pages} {042317} (\bibinfo {year} {2015})}\BibitemShut
  {NoStop}%
\bibitem [{\citenamefont {Quintino}\ \emph {et~al.}(2015)\citenamefont
  {Quintino}, \citenamefont {V\'ertesi}, \citenamefont {Cavalcanti},
  \citenamefont {Augusiak}, \citenamefont {Demianowicz}, \citenamefont
  {Ac\'{\i}n},\ and\ \citenamefont {Brunner}}]{QVC+15}%
  \BibitemOpen
  \bibfield  {author} {\bibinfo {author} {\bibfnamefont {Marco~T\'ulio}\
  \bibnamefont {Quintino}}, \bibinfo {author} {\bibfnamefont {Tam\'as}\
  \bibnamefont {V\'ertesi}}, \bibinfo {author} {\bibfnamefont {Daniel}\
  \bibnamefont {Cavalcanti}}, \bibinfo {author} {\bibfnamefont {Remigiusz}\
  \bibnamefont {Augusiak}}, \bibinfo {author} {\bibfnamefont {Maciej}\
  \bibnamefont {Demianowicz}}, \bibinfo {author} {\bibfnamefont {Antonio}\
  \bibnamefont {Ac\'{\i}n}}, \ and\ \bibinfo {author} {\bibfnamefont {Nicolas}\
  \bibnamefont {Brunner}},\ }\bibfield  {title} {\enquote {\bibinfo {title}
  {Inequivalence of entanglement, steering, and bell nonlocality for general
  measurements},}\ }\href {\doibase 10.1103/PhysRevA.92.032107} {\bibfield
  {journal} {\bibinfo  {journal} {Phys. Rev. A}\ }\textbf {\bibinfo {volume}
  {92}},\ \bibinfo {pages} {032107} (\bibinfo {year} {2015})}\BibitemShut
  {NoStop}%
\bibitem [{\citenamefont {Jones}\ \emph {et~al.}(2007)\citenamefont {Jones},
  \citenamefont {Wiseman},\ and\ \citenamefont {Doherty}}]{JWD07}%
  \BibitemOpen
  \bibfield  {author} {\bibinfo {author} {\bibfnamefont {S.~J.}\ \bibnamefont
  {Jones}}, \bibinfo {author} {\bibfnamefont {H.~M.}\ \bibnamefont {Wiseman}},
  \ and\ \bibinfo {author} {\bibfnamefont {A.~C.}\ \bibnamefont {Doherty}},\
  }\bibfield  {title} {\enquote {\bibinfo {title} {Entanglement,
  einstein-podolsky-rosen correlations, bell nonlocality, and steering},}\
  }\href {\doibase 10.1103/PhysRevA.76.052116} {\bibfield  {journal} {\bibinfo
  {journal} {Phys. Rev. A}\ }\textbf {\bibinfo {volume} {76}},\ \bibinfo
  {pages} {052116} (\bibinfo {year} {2007})}\BibitemShut {NoStop}%
\bibitem [{\citenamefont {Cavalcanti}\ \emph {et~al.}(2009)\citenamefont
  {Cavalcanti}, \citenamefont {Jones}, \citenamefont {Wiseman},\ and\
  \citenamefont {Reid}}]{CJW+09}%
  \BibitemOpen
  \bibfield  {author} {\bibinfo {author} {\bibfnamefont {E.~G.}\ \bibnamefont
  {Cavalcanti}}, \bibinfo {author} {\bibfnamefont {S.~J.}\ \bibnamefont
  {Jones}}, \bibinfo {author} {\bibfnamefont {H.~M.}\ \bibnamefont {Wiseman}},
  \ and\ \bibinfo {author} {\bibfnamefont {M.~D.}\ \bibnamefont {Reid}},\
  }\bibfield  {title} {\enquote {\bibinfo {title} {Experimental criteria for
  steering and the einstein-podolsky-rosen paradox},}\ }\href {\doibase
  10.1103/PhysRevA.80.032112} {\bibfield  {journal} {\bibinfo  {journal} {Phys.
  Rev. A}\ }\textbf {\bibinfo {volume} {80}},\ \bibinfo {pages} {032112}
  (\bibinfo {year} {2009})}\BibitemShut {NoStop}%
\bibitem [{\citenamefont {Zhu}\ \emph {et~al.}(2016)\citenamefont {Zhu},
  \citenamefont {Hayashi},\ and\ \citenamefont {Chen}}]{ZHC16}%
  \BibitemOpen
  \bibfield  {author} {\bibinfo {author} {\bibfnamefont {Huangjun}\
  \bibnamefont {Zhu}}, \bibinfo {author} {\bibfnamefont {Masahito}\
  \bibnamefont {Hayashi}}, \ and\ \bibinfo {author} {\bibfnamefont {Lin}\
  \bibnamefont {Chen}},\ }\bibfield  {title} {\enquote {\bibinfo {title}
  {Universal steering criteria},}\ }\href {\doibase
  10.1103/PhysRevLett.116.070403} {\bibfield  {journal} {\bibinfo  {journal}
  {Phys. Rev. Lett.}\ }\textbf {\bibinfo {volume} {116}},\ \bibinfo {pages}
  {070403} (\bibinfo {year} {2016})}\BibitemShut {NoStop}%
\bibitem [{\citenamefont {Branciard}\ \emph {et~al.}(2012)\citenamefont
  {Branciard}, \citenamefont {Cavalcanti}, \citenamefont {Walborn},
  \citenamefont {Scarani},\ and\ \citenamefont {Wiseman}}]{BCW+12}%
  \BibitemOpen
  \bibfield  {author} {\bibinfo {author} {\bibfnamefont {Cyril}\ \bibnamefont
  {Branciard}}, \bibinfo {author} {\bibfnamefont {Eric~G.}\ \bibnamefont
  {Cavalcanti}}, \bibinfo {author} {\bibfnamefont {Stephen~P.}\ \bibnamefont
  {Walborn}}, \bibinfo {author} {\bibfnamefont {Valerio}\ \bibnamefont
  {Scarani}}, \ and\ \bibinfo {author} {\bibfnamefont {Howard~M.}\ \bibnamefont
  {Wiseman}},\ }\bibfield  {title} {\enquote {\bibinfo {title} {One-sided
  device-independent quantum key distribution: Security, feasibility, and the
  connection with steering},}\ }\href {\doibase 10.1103/PhysRevA.85.010301}
  {\bibfield  {journal} {\bibinfo  {journal} {Phys. Rev. A}\ }\textbf {\bibinfo
  {volume} {85}},\ \bibinfo {pages} {010301 (R)} (\bibinfo {year}
  {2012})}\BibitemShut {NoStop}%
\bibitem [{\citenamefont {Chen}\ \emph {et~al.}(2013)\citenamefont {Chen},
  \citenamefont {Ye}, \citenamefont {Wu}, \citenamefont {Su}, \citenamefont
  {Cabello}, \citenamefont {Kwek},\ and\ \citenamefont {Oh}}]{CYW+13}%
  \BibitemOpen
  \bibfield  {author} {\bibinfo {author} {\bibfnamefont {J. L.}\ \bibnamefont
  {Chen}}, \bibinfo {author} {\bibfnamefont {X. J.}\ \bibnamefont {Ye}},
  \bibinfo {author} {\bibfnamefont {C. F.}\ \bibnamefont {Wu}}, \bibinfo
  {author} {\bibfnamefont {H. Y.}\ \bibnamefont {Su}}, \bibinfo {author}
  {\bibfnamefont {A.}~\bibnamefont {Cabello}}, \bibinfo {author} {\bibfnamefont
  {L. C.}\ \bibnamefont {Kwek}}, \ and\ \bibinfo {author} {\bibfnamefont
  {C. H.}\ \bibnamefont {Oh}},\ }\bibfield  {title} {\enquote {\bibinfo
  {title} {All-versus-nothing proof of einstein-podolsky-rosen steering},}\
  }\href {\doibase doi:10.1038/srep02143} {\bibfield  {journal} {\bibinfo
  {journal} {Scientific Reports}\ }\textbf {\bibinfo {volume} {3}},\ \bibinfo
  {pages} {2143} (\bibinfo {year} {2013})}\BibitemShut {NoStop}%
\bibitem [{\citenamefont {Bowles}\ \emph {et~al.}(2014)\citenamefont {Bowles},
  \citenamefont {V\'ertesi}, \citenamefont {Quintino},\ and\ \citenamefont
  {Brunner}}]{BVQ+14}%
  \BibitemOpen
  \bibfield  {author} {\bibinfo {author} {\bibfnamefont {Joseph}\ \bibnamefont
  {Bowles}}, \bibinfo {author} {\bibfnamefont {Tam\'as}\ \bibnamefont
  {V\'ertesi}}, \bibinfo {author} {\bibfnamefont {Marco~T\'ulio}\ \bibnamefont
  {Quintino}}, \ and\ \bibinfo {author} {\bibfnamefont {Nicolas}\ \bibnamefont
  {Brunner}},\ }\bibfield  {title} {\enquote {\bibinfo {title} {One-way
  einstein-podolsky-rosen steering},}\ }\href {\doibase
  10.1103/PhysRevLett.112.200402} {\bibfield  {journal} {\bibinfo  {journal}
  {Phys. Rev. Lett.}\ }\textbf {\bibinfo {volume} {112}},\ \bibinfo {pages}
  {200402} (\bibinfo {year} {2014})}\BibitemShut {NoStop}%
\bibitem [{\citenamefont {Kogias}\ \emph {et~al.}(2015)\citenamefont {Kogias},
  \citenamefont {Skrzypczyk}, \citenamefont {Cavalcanti}, \citenamefont
  {Ac\'{\i}n},\ and\ \citenamefont {Adesso}}]{KSC+15}%
  \BibitemOpen
  \bibfield  {author} {\bibinfo {author} {\bibfnamefont {Ioannis}\ \bibnamefont
  {Kogias}}, \bibinfo {author} {\bibfnamefont {Paul}\ \bibnamefont
  {Skrzypczyk}}, \bibinfo {author} {\bibfnamefont {Daniel}\ \bibnamefont
  {Cavalcanti}}, \bibinfo {author} {\bibfnamefont {Antonio}\ \bibnamefont
  {Ac\'{\i}n}}, \ and\ \bibinfo {author} {\bibfnamefont {Gerardo}\ \bibnamefont
  {Adesso}},\ }\bibfield  {title} {\enquote {\bibinfo {title} {Hierarchy of
  steering criteria based on moments for all bipartite quantum systems},}\
  }\href {\doibase 10.1103/PhysRevLett.115.210401} {\bibfield  {journal}
  {\bibinfo  {journal} {Phys. Rev. Lett.}\ }\textbf {\bibinfo {volume} {115}},\
  \bibinfo {pages} {210401} (\bibinfo {year} {2015})}\BibitemShut {NoStop}%
\bibitem [{\citenamefont {Skrzypczyk}\ \emph {et~al.}(2014)\citenamefont
  {Skrzypczyk}, \citenamefont {Navascu\'es},\ and\ \citenamefont
  {Cavalcanti}}]{SNC14}%
  \BibitemOpen
  \bibfield  {author} {\bibinfo {author} {\bibfnamefont {Paul}\ \bibnamefont
  {Skrzypczyk}}, \bibinfo {author} {\bibfnamefont {Miguel}\ \bibnamefont
  {Navascu\'es}}, \ and\ \bibinfo {author} {\bibfnamefont {Daniel}\
  \bibnamefont {Cavalcanti}},\ }\bibfield  {title} {\enquote {\bibinfo {title}
  {Quantifying einstein-podolsky-rosen steering},}\ }\href {\doibase
  10.1103/PhysRevLett.112.180404} {\bibfield  {journal} {\bibinfo  {journal}
  {Phys. Rev. Lett.}\ }\textbf {\bibinfo {volume} {112}},\ \bibinfo {pages}
  {180404} (\bibinfo {year} {2014})}\BibitemShut {NoStop}%
\bibitem [{\citenamefont {Gallego}\ and\ \citenamefont {Aolita}(2015)}]{GA15}%
  \BibitemOpen
  \bibfield  {author} {\bibinfo {author} {\bibfnamefont {Rodrigo}\ \bibnamefont
  {Gallego}}\ and\ \bibinfo {author} {\bibfnamefont {Leandro}\ \bibnamefont
  {Aolita}},\ }\bibfield  {title} {\enquote {\bibinfo {title} {Resource theory
  of steering},}\ }\href {\doibase 10.1103/PhysRevX.5.041008} {\bibfield
  {journal} {\bibinfo  {journal} {Phys. Rev. X}\ }\textbf {\bibinfo {volume}
  {5}},\ \bibinfo {pages} {041008} (\bibinfo {year} {2015})}\BibitemShut
  {NoStop}%
\bibitem [{\citenamefont {Chen}\ \emph {et~al.}(2016)\citenamefont {Chen},
  \citenamefont {Budroni}, \citenamefont {Liang},\ and\ \citenamefont
  {Chen}}]{CBL+16}%
  \BibitemOpen
  \bibfield  {author} {\bibinfo {author} {\bibfnamefont {Shin-Liang}\
  \bibnamefont {Chen}}, \bibinfo {author} {\bibfnamefont {Costantino}\
  \bibnamefont {Budroni}}, \bibinfo {author} {\bibfnamefont {Yeong-Cherng}\
  \bibnamefont {Liang}}, \ and\ \bibinfo {author} {\bibfnamefont {Yueh-Nan}\
  \bibnamefont {Chen}},\ }\bibfield  {title} {\enquote {\bibinfo {title}
  {Natural framework for device-independent quantification of quantum
  steerability, measurement incompatibility, and self-testing},}\ }\href
  {\doibase 10.1103/PhysRevLett.116.240401} {\bibfield  {journal} {\bibinfo
  {journal} {Phys. Rev. Lett.}\ }\textbf {\bibinfo {volume} {116}},\ \bibinfo
  {pages} {240401} (\bibinfo {year} {2016})}\BibitemShut {NoStop}%
\bibitem [{\citenamefont {Fine}(1982)}]{Fin82}%
  \BibitemOpen
  \bibfield  {author} {\bibinfo {author} {\bibfnamefont {Arthur}\ \bibnamefont
  {Fine}},\ }\bibfield  {title} {\enquote {\bibinfo {title} {Hidden variables,
  joint probability, and the bell inequalities},}\ }\href {\doibase
  10.1103/PhysRevLett.48.291} {\bibfield  {journal} {\bibinfo  {journal} {Phys.
  Rev. Lett.}\ }\textbf {\bibinfo {volume} {48}},\ \bibinfo {pages} {291--295}
  (\bibinfo {year} {1982})}\BibitemShut {NoStop}%
\bibitem [{\citenamefont {Popescu}\ and\ \citenamefont
  {Rohrlich}(1994)}]{PR94}%
  \BibitemOpen
  \bibfield  {author} {\bibinfo {author} {\bibfnamefont {Sandu}\ \bibnamefont
  {Popescu}}\ and\ \bibinfo {author} {\bibfnamefont {Daniel}\ \bibnamefont
  {Rohrlich}},\ }\bibfield  {title} {\enquote {\bibinfo {title} {Quantum
  nonlocality as an axiom},}\ }\href {\doibase 10.1007/BF02058098} {\bibfield
  {journal} {\bibinfo  {journal} {Found. Phys.}\ }\textbf {\bibinfo {volume}
  {24}},\ \bibinfo {pages} {379--385} (\bibinfo {year} {1994})}\BibitemShut
  {NoStop}%
\bibitem [{\citenamefont {Barrett}\ \emph {et~al.}(2005)\citenamefont
  {Barrett}, \citenamefont {Linden}, \citenamefont {Massar}, \citenamefont
  {Pironio}, \citenamefont {Popescu},\ and\ \citenamefont {Roberts}}]{BLM+05}%
  \BibitemOpen
  \bibfield  {author} {\bibinfo {author} {\bibfnamefont {Jonathan}\
  \bibnamefont {Barrett}}, \bibinfo {author} {\bibfnamefont {Noah}\
  \bibnamefont {Linden}}, \bibinfo {author} {\bibfnamefont {Serge}\
  \bibnamefont {Massar}}, \bibinfo {author} {\bibfnamefont {Stefano}\
  \bibnamefont {Pironio}}, \bibinfo {author} {\bibfnamefont {Sandu}\
  \bibnamefont {Popescu}}, \ and\ \bibinfo {author} {\bibfnamefont {David}\
  \bibnamefont {Roberts}},\ }\bibfield  {title} {\enquote {\bibinfo {title}
  {Nonlocal correlations as an information-theoretic resource},}\ }\href
  {\doibase 10.1103/PhysRevA.71.022101} {\bibfield  {journal} {\bibinfo
  {journal} {Phys. Rev. A}\ }\textbf {\bibinfo {volume} {71}},\ \bibinfo
  {pages} {022101} (\bibinfo {year} {2005})}\BibitemShut {NoStop}%
\bibitem [{\citenamefont {Cavalcanti}\ \emph {et~al.}(2015)\citenamefont
  {Cavalcanti}, \citenamefont {Foster}, \citenamefont {Fuwa},\ and\
  \citenamefont {Wiseman}}]{CFF+15}%
  \BibitemOpen
  \bibfield  {author} {\bibinfo {author} {\bibfnamefont {Eric~G.}\ \bibnamefont
  {Cavalcanti}}, \bibinfo {author} {\bibfnamefont {Christopher~J.}\
  \bibnamefont {Foster}}, \bibinfo {author} {\bibfnamefont {Maria}\
  \bibnamefont {Fuwa}}, \ and\ \bibinfo {author} {\bibfnamefont {Howard~M.}\
  \bibnamefont {Wiseman}},\ }\bibfield  {title} {\enquote {\bibinfo {title}
  {Analog of the clauser--horne--shimony--holt inequality for steering},}\
  }\href {\doibase 10.1364/JOSAB.32.000A74} {\bibfield  {journal} {\bibinfo
  {journal} {J. Opt. Soc. Am. B}\ }\textbf {\bibinfo {volume} {32}},\ \bibinfo
  {pages} {A74--A81} (\bibinfo {year} {2015})}\BibitemShut {NoStop}%
\bibitem [{\citenamefont {Clauser}\ \emph {et~al.}(1969)\citenamefont
  {Clauser}, \citenamefont {Horne}, \citenamefont {Shimony},\ and\
  \citenamefont {Holt}}]{CHS+69}%
  \BibitemOpen
  \bibfield  {author} {\bibinfo {author} {\bibfnamefont {John~F.}\ \bibnamefont
  {Clauser}}, \bibinfo {author} {\bibfnamefont {Michael~A.}\ \bibnamefont
  {Horne}}, \bibinfo {author} {\bibfnamefont {Abner}\ \bibnamefont {Shimony}},
  \ and\ \bibinfo {author} {\bibfnamefont {Richard~A.}\ \bibnamefont {Holt}},\
  }\bibfield  {title} {\enquote {\bibinfo {title} {Proposed experiment to test
  local hidden-variable theories},}\ }\href {\doibase
  10.1103/PhysRevLett.23.880} {\bibfield  {journal} {\bibinfo  {journal} {Phys.
  Rev. Lett.}\ }\textbf {\bibinfo {volume} {23}},\ \bibinfo {pages} {880--884}
  (\bibinfo {year} {1969})}\BibitemShut {NoStop}%
\bibitem [{\citenamefont {Brunner}\ \emph {et~al.}(2011)\citenamefont
  {Brunner}, \citenamefont {Cavalcanti}, \citenamefont {Salles},\ and\
  \citenamefont {Skrzypczyk}}]{BCS+11}%
  \BibitemOpen
  \bibfield  {author} {\bibinfo {author} {\bibfnamefont {Nicolas}\ \bibnamefont
  {Brunner}}, \bibinfo {author} {\bibfnamefont {Daniel}\ \bibnamefont
  {Cavalcanti}}, \bibinfo {author} {\bibfnamefont {Alejo}\ \bibnamefont
  {Salles}}, \ and\ \bibinfo {author} {\bibfnamefont {Paul}\ \bibnamefont
  {Skrzypczyk}},\ }\bibfield  {title} {\enquote {\bibinfo {title} {Bound
  nonlocality and activation},}\ }\href {\doibase
  10.1103/PhysRevLett.106.020402} {\bibfield  {journal} {\bibinfo  {journal}
  {Phys. Rev. Lett.}\ }\textbf {\bibinfo {volume} {106}},\ \bibinfo {pages}
  {020402} (\bibinfo {year} {2011})}\BibitemShut {NoStop}%
\bibitem [{\citenamefont {Elitzur}\ \emph {et~al.}(1992)\citenamefont
  {Elitzur}, \citenamefont {Popescu},\ and\ \citenamefont {Rohrlich}}]{EPR92}%
  \BibitemOpen
  \bibfield  {author} {\bibinfo {author} {\bibfnamefont {Avshalom~C.}\
  \bibnamefont {Elitzur}}, \bibinfo {author} {\bibfnamefont {Sandu}\
  \bibnamefont {Popescu}}, \ and\ \bibinfo {author} {\bibfnamefont {Daniel}\
  \bibnamefont {Rohrlich}},\ }\bibfield  {title} {\enquote {\bibinfo {title}
  {Quantum nonlocality for each pair in an ensemble},}\ }\href {\doibase
  http://dx.doi.org/10.1016/0375-9601(92)90952-I} {\bibfield  {journal}
  {\bibinfo  {journal} {Phys. Lett. A}\ }\textbf {\bibinfo {volume} {162}},\
  \bibinfo {pages} {25 -- 28} (\bibinfo {year} {1992})}\BibitemShut {NoStop}%
\bibitem [{\citenamefont {Pusey}(2013)}]{Pus13}%
  \BibitemOpen
  \bibfield  {author} {\bibinfo {author} {\bibfnamefont {Matthew~F.}\
  \bibnamefont {Pusey}},\ }\bibfield  {title} {\enquote {\bibinfo {title}
  {Negativity and steering: A stronger peres conjecture},}\ }\href {\doibase
  10.1103/PhysRevA.88.032313} {\bibfield  {journal} {\bibinfo  {journal} {Phys.
  Rev. A}\ }\textbf {\bibinfo {volume} {88}},\ \bibinfo {pages} {032313}
  (\bibinfo {year} {2013})}\BibitemShut {NoStop}%
\bibitem [{\citenamefont {Cavalcanti}\ and\ \citenamefont
  {Skrzypczyk}(2017)}]{CS17}%
  \BibitemOpen
  \bibfield  {author} {\bibinfo {author} {\bibfnamefont {D}~\bibnamefont
  {Cavalcanti}}\ and\ \bibinfo {author} {\bibfnamefont {P}~\bibnamefont
  {Skrzypczyk}},\ }\bibfield  {title} {\enquote {\bibinfo {title} {Quantum
  steering: a review with focus on semidefinite programming},}\ }\href
  {http://stacks.iop.org/0034-4885/80/i=2/a=024001} {\bibfield  {journal}
  {\bibinfo  {journal} {Reports on Progress in Physics}\ }\textbf {\bibinfo
  {volume} {80}},\ \bibinfo {pages} {024001} (\bibinfo {year}
  {2017})}\BibitemShut {NoStop}%
\bibitem [{\citenamefont {Hsieh}\ \emph {et~al.}(2016)\citenamefont {Hsieh},
  \citenamefont {Liang},\ and\ \citenamefont {Lee}}]{HLL16}%
  \BibitemOpen
  \bibfield  {author} {\bibinfo {author} {\bibfnamefont {Chung-Yun}\
  \bibnamefont {Hsieh}}, \bibinfo {author} {\bibfnamefont {Yeong-Cherng}\
  \bibnamefont {Liang}}, \ and\ \bibinfo {author} {\bibfnamefont {Ray-Kuang}\
  \bibnamefont {Lee}},\ }\bibfield  {title} {\enquote {\bibinfo {title}
  {Quantum steerability: Characterization, quantification, superactivation, and
  unbounded amplification},}\ }\href {\doibase 10.1103/PhysRevA.94.062120}
  {\bibfield  {journal} {\bibinfo  {journal} {Phys. Rev. A}\ }\textbf {\bibinfo
  {volume} {94}},\ \bibinfo {pages} {062120} (\bibinfo {year}
  {2016})}\BibitemShut {NoStop}%
\bibitem [{\citenamefont {Donohue}\ and\ \citenamefont {Wolfe}(2015)}]{DW15}%
  \BibitemOpen
  \bibfield  {author} {\bibinfo {author} {\bibfnamefont {John~Matthew}\
  \bibnamefont {Donohue}}\ and\ \bibinfo {author} {\bibfnamefont {Elie}\
  \bibnamefont {Wolfe}},\ }\bibfield  {title} {\enquote {\bibinfo {title}
  {Identifying nonconvexity in the sets of limited-dimension quantum
  correlations},}\ }\href {\doibase 10.1103/PhysRevA.92.062120} {\bibfield
  {journal} {\bibinfo  {journal} {Phys. Rev. A}\ }\textbf {\bibinfo {volume}
  {92}},\ \bibinfo {pages} {062120} (\bibinfo {year} {2015})}\BibitemShut
  {NoStop}%
\bibitem [{\citenamefont {Ac\'{\i}n}\ \emph {et~al.}(2006)\citenamefont
  {Ac\'{\i}n}, \citenamefont {Gisin},\ and\ \citenamefont {Masanes}}]{AGM06}%
  \BibitemOpen
  \bibfield  {author} {\bibinfo {author} {\bibfnamefont {Antonio}\ \bibnamefont
  {Ac\'{\i}n}}, \bibinfo {author} {\bibfnamefont {Nicolas}\ \bibnamefont
  {Gisin}}, \ and\ \bibinfo {author} {\bibfnamefont {Lluis}\ \bibnamefont
  {Masanes}},\ }\bibfield  {title} {\enquote {\bibinfo {title} {From bell's
  theorem to secure quantum key distribution},}\ }\href {\doibase
  10.1103/PhysRevLett.97.120405} {\bibfield  {journal} {\bibinfo  {journal}
  {Phys. Rev. Lett.}\ }\textbf {\bibinfo {volume} {97}},\ \bibinfo {pages}
  {120405} (\bibinfo {year} {2006})}\BibitemShut {NoStop}%
\bibitem [{\citenamefont {Werner}(1989)}]{Wer89}%
  \BibitemOpen
  \bibfield  {author} {\bibinfo {author} {\bibfnamefont {Reinhard~F.}\
  \bibnamefont {Werner}},\ }\bibfield  {title} {\enquote {\bibinfo {title}
  {Quantum states with einstein-podolsky-rosen correlations admitting a
  hidden-variable model},}\ }\href {\doibase 10.1103/PhysRevA.40.4277}
  {\bibfield  {journal} {\bibinfo  {journal} {Phys. Rev. A}\ }\textbf {\bibinfo
  {volume} {40}},\ \bibinfo {pages} {4277--4281} (\bibinfo {year}
  {1989})}\BibitemShut {NoStop}%
\bibitem [{\citenamefont {de~Vicente}(2014)}]{Vicente}%
  \BibitemOpen
  \bibfield  {author} {\bibinfo {author} {\bibfnamefont {Julio~I}\ \bibnamefont
  {de~Vicente}},\ }\bibfield  {title} {\enquote {\bibinfo {title} {On
  nonlocality as a resource theory and nonlocality measures},}\ }\href
  {http://stacks.iop.org/1751-8121/47/i=42/a=424017} {\bibfield  {journal}
  {\bibinfo  {journal} {Journal of Physics A: Mathematical and Theoretical}\
  }\textbf {\bibinfo {volume} {47}},\ \bibinfo {pages} {424017} (\bibinfo
  {year} {2014})}\BibitemShut {NoStop}%
\bibitem [{\citenamefont {James}\ \emph {et~al.}(2001)\citenamefont {James},
  \citenamefont {Kwiat}, \citenamefont {Munro},\ and\ \citenamefont
  {White}}]{JKM+01}%
  \BibitemOpen
  \bibfield  {author} {\bibinfo {author} {\bibfnamefont {Daniel F.~V.}\
  \bibnamefont {James}}, \bibinfo {author} {\bibfnamefont {Paul~G.}\
  \bibnamefont {Kwiat}}, \bibinfo {author} {\bibfnamefont {William~J.}\
  \bibnamefont {Munro}}, \ and\ \bibinfo {author} {\bibfnamefont {Andrew~G.}\
  \bibnamefont {White}},\ }\bibfield  {title} {\enquote {\bibinfo {title}
  {Measurement of qubits},}\ }\href {\doibase 10.1103/PhysRevA.64.052312}
  {\bibfield  {journal} {\bibinfo  {journal} {Phys. Rev. A}\ }\textbf {\bibinfo
  {volume} {64}},\ \bibinfo {pages} {052312} (\bibinfo {year}
  {2001})}\BibitemShut {NoStop}%
\end{thebibliography}%

\end{document}